\documentclass[lettersize,journal]{IEEEtran}
\usepackage{amsthm}
\usepackage{amsmath,amsfonts}
\usepackage{algorithmic}
\usepackage{algorithm}
\usepackage{amssymb}
\usepackage{array}
\usepackage[caption=false,font=normalsize,labelfont=sf,textfont=sf]{subfig}
\usepackage{textcomp}
\usepackage{stfloats}
\usepackage{url}
\usepackage{verbatim}
\usepackage{graphicx}
\usepackage{cite}
\usepackage{cases}
\usepackage{hyperref}
\usepackage{bm}
\usepackage{color}
\usepackage{gensymb}
\usepackage{multirow}

\hypersetup{colorlinks=true,
	linkcolor=blue,
	citecolor=blue,      
	urlcolor=black,
}

\newtheoremstyle{mylemma}
{}{}                 
{\normalfont}        
{}                   
{\bfseries}         
{.}                  
{ }                  
{\thmname{#1}\thmnumber{ #2}\thmnote{ (#3)}} 

\theoremstyle{mylemma}

\newtheorem{remark}{Remark}

\begin{document}

\title{Beam-Brainstorm: A Generative Site-Specific Beamforming Approach}

\author{Zihao Zhou, Zhaolin Wang,~\IEEEmembership{Member,~IEEE}, and Yuanwei Liu,~\IEEEmembership{Fellow,~IEEE}
\thanks{The authors are with the Department of Electrical and Electronic Engineering, The University of Hong Kong, Hong Kong (e-mail: eezihaozhou@connect.hku.hk,zhaolin.wang@hku.hk,yuanwei@hku.hk)}}

\maketitle

\begin{abstract}
Accurately understanding the propagation environment is a fundamental challenge in site-specific beamforming (SSBF). This paper proposes a novel generative SSBF (GenSSBF) solution, which represents a paradigm shift from conventional unstructured prediction to joint-structure modeling. First, considering the fundamental differences between beam generation and conventional image synthesis, a unified GenSSBF framework is proposed, which includes a site profile, a wireless prompting module, and a generator. Second, a beam-brainstorm (BBS) solution is proposed as an instantiation of this GenSSBF framework. Specifically, the site profile is configured by transforming channel data from spatial domain to a reversible latent space via discrete Fourier transform (DFT). To facilitate practical deployment, the wireless prompt is constructed from the reference signal received power (RSRP) measured using a small number of DFT-beams. Finally, the generator is developed using a customized conditional diffusion model. Rather than relying on a meticulously designed global codebook, BBS directly generates diverse and high-fidelity user-specific beams guided by the wireless prompts. Simulation results on accurate ray-tracing datasets demonstrate that BBS can achieve near-optimal beamforming gain while drastically reducing the beam sweeping overhead, even in low signal-to-noise ratio (SNR) environments. 
\end{abstract}

\begin{IEEEkeywords}
Beamforming, generative diffusion models, generative AI, 6G.
\end{IEEEkeywords}
\vspace{-1.2em}
\section{Introduction}
\IEEEPARstart{D}{ecades} of advancement have cemented multi-antenna technology as a cornerstone of modern wireless communication systems\cite{mietzner2009multiple}. An important enabler in this evolution is beamforming, which unlocks extraordinary performance gains through spatial diversity and multiplexing\cite{wong2021fluid}. This is particularly critical for networks operating at millimeter wave (mmWave) or terahertz (THz) frequencies, where highly directional beamforming is indispensable to compensate the substantial propagation and penetration losses\cite{xiao2017millimeter, kutty2016beamforming, kim2024fast}. Theoretically, achieving optimal beamforming requires accurate and timely and perfect channel state information (CSI). In practice, however, given the complex wireless environments, the growing dimensionality of massive arrays, and user mobility, the real-time acquisition of perfect CSI is infeasible.

The beamforming approaches in the 5th generation new radio (5G NR) are primarily follows two paradigms\cite{dahlman20185g}: eigen based, and beam sweeping based. For the former, the system typically follows the ``estimate-and-optimize'' procedure. Specifically, the CSI should be first estimated through the transmission of pilot signals based on uplink-downlink reciprocity, and then the base station (BS) adopts complex optimization algorithms to compute the beamformers based on the obtained CSI\cite{ngo2017cell, guo2021deep, li2021network}. However, CSI estimation requires substantial pilot resources and intensive baseband processing\cite{shen2017high}. This overhead becomes especially acute in systems with a colossal number of antennas. By contrast, the beam sweeping based methods are more commonly used, which involve sequential probing, measuring and reporting\cite{heng2021six}. This paradigm relies on a pre-defined codebook whose beams can cover all possible quantized directions (e.g., discrete Fourier transform (DFT) or oversampled-DFT (O-DFT) codebook\cite{abdallah2025explainable}). Specifically, in downlink, the BS employs synchronization signal blocks (SSBs) to exhaustively sweep the probing beams of its codebook. The user equipment (UE) monitors the received signal power, and then reports the measurements to the BS. Based on this feedback from UE, the best beam can be determined by the BS. However, when the number of narrow beams in the codebook surges, such exhaustive search will lead to extremely high beam sweeping overhead. Therefore, hierarchical search methods were proposed\cite{he2015suboptimal, xiao2016hierarchical, qi2020hierarchical} where the BS uses SSBs to scan wider beams first, and progress to narrower child beams using more flexible CSI reference signals (CSI-RSs).

It is noticed that the codebooks employed in the aforementioned beam sweeping based methods are \emph{site-agnostic}, that is, the same codebook is applied universally across different sites. However, the wireless propagation environment is inherently \emph{site-specific}, as it is shaped by local geometric features, user distributions, and hardware adopted, all of which vary from one deployment site to another. Therefore, the traditional site-agnostic codebooks suffer from the following inherent limitations:
\begin{itemize}
	\item \textbf{Environmental unawareness}: The beams in codebook strive to cover all possible angular directions, but for some specific sites (e.g., hallways, hotspots), some of the directions may never be used\cite{alrabeia2022neural}. This results in prolonged beam sweeping time and unnecessary overhead.
	\item \textbf{Performance bottleneck}: The beams in DFT or O-DFT codebooks are typically single-lobe, which limits the beamforming gain especially in non line-of-sight (NLoS) scenarios. Furthermore, these on-grid methods suffer from ``grid mismatch'' problem\cite{abdallah2025explainable}, where the predefined beams may not align perfectly with the actual channel, leading to suboptimal performance.
	\item \textbf{Limited scalability}: With larger antenna arrays (e.g., extremely large-scale MIMO (XL-MIMO) with several hundreds of even thousands of antennas\cite{liu2023near}), the size of these site-agnostic codebooks will also surge significantly. Conventional beam search optimization can only offer marginal gains.
\end{itemize}

Therefore, beamforming techniques towards 6G must evolve to incorporate \emph{spatial intelligence}, a paradigm referred to as \emph{site-specific beamforming} (SSBF)\cite{heng2024site}. Recent years have witnessed a growing focus on deep learning (DL)-assisted SSBF. In \cite{alrabeia2022neural}, a site-specific probing codebook was learnt offline by a complex-valued neural network (NN) with channel vectors as input. The beamforming weights of the analog phase shifters were directly represented by the neuron weights. Following the same idea of directly characterizing the codebook by complex NN, the authors of \cite{heng2022learning} added a new prediction layer to predict the optimal narrow beam index in DFT codebook, based on the power measurements obtained from sweeping the learnt probing codebook. In \cite{yang2024hierarchical}, the framework of \cite{heng2022learning} was adapted to hierarchical beam alignment, where at each layer, the probing codebook and the next-layer selector were jointly trained. Rather than using channel data via ray-tracing, the authors of \cite{kwak2024site} proposed to adopt power measurements of the received signal as the dataset to train an end-to-end codebook learning and narrow beam index prediction framework. In \cite{chen2023computer}, a computer vision (CV)-based site-specific codebook design method was proposed, where the codebook was learnt from the images captured at the BS for LoS users, and from features extracted from the 3D point cloud for NLoS users. The authors of \cite{luo2025digital} proposed using a site-specific digital twin to generate synthetic channel data, which were utilized for codebook learning.

Although the aforementioned approaches\cite{alrabeia2022neural,heng2022learning,yang2024hierarchical,kwak2024site,chen2023computer,luo2025digital} learn more ``meaningful'' codebooks that capture site-specific information, the communication beams are still on-grid. Since the size of the codebook is limited, the overall beamforming gain of the system remains constrained, regardless of whether the beam is selected from the learnt codebook or the traditional site-agnostic one. Furthermore, the issue of ``grid mismatch'' mentioned above still remains. In \cite{heng2024grid}, a grid-free site-specific beam alignment framework was proposed where the probing codebooks and the multi-layer perceptron (MLP)-based beam synthesizers were jointly trained. Similarly, the application of a MLP for beam synthesis is seen in \cite{abdallah2025explainable} for hybrid beamforming and in \cite{li2025site} for joint transmit- and receive beamforming in full-duplex system. However, due to the complexity of multi-path propagation environments, different users may have similar coarse CSI measurements (the input to these discriminative models), while their corresponding optimal beamforming vectors can be totally different. This is a phenomenon termed \textit{multimodality} in SSBF. In such cases, conventional discriminative regression based on mean squared error (MSE) loss tends to average across these multiple plausible beams, leading to significant performance degradation. Furthermore, these methods essentially belong to ``\textit{unstructured prediction}''\cite{torralba2024foundations} which assumes conditional independence among all elements of the beamforming vector given the input. This assumption, however, fails to account for the fact that these elements are highly correlated and collectively determine whether the signals combine constructively or destructively at the receiver\cite{masouros2015exploiting}.

In contrast to the discriminative models, generative artificial intelligence (GenAI) models fall under the category of ``\textit{structured prediction}'', and can learn the multimodal distribution, offering a transformative potential that extends beyond the traditional boundaries of discriminative models\cite{du2024enhancing, xu2025mobile}. Recent work has also demonstrated the significant role of GenAI (e.g., diffusion models\cite{ho2020denoising, song2020denoising}) in wireless communications. In \cite{lee2025generating}, given a small number of true channel matrices, the conditional denoising diffusion implicit model (cDDIM) was employed to create more augmented channel samples. Similarly, the diffusion model-assisted channel estimation, extrapolation, and feedback were explored in MIMO systems\cite{xu2025generative}. The authors of \cite{kim2023diffusion} treated diffusion model as a differentiable channel generator in a end-to-end (E2E) wireless communication system, which enables the design/training of other modules (e.g., encoder/decoder) under channel distribution with high-fidelity. In \cite{wang2025energy}, diffusion model was adopted as an actor network in deep reinforcement learning (DRL) framework to solve a resource allocation problem under the context of rate-splitting multiple access (RSMA) based low-altitude mobile edge computing (MEG). Likewise, following the paradigm of ``diffusion model as an actor in DRL'', the authors of \cite{zhang2025enhanced} investigated the secure beamforming in intelligent reflecting surface (RIS)-assisted Internet of Things (IoT) communications. However, the channel matrices, transmission/eavesdropping rates, as well as the historical actions are required to make next decision, which will lead to significant signaling overhead and latency.

For practical SSBF tasks, which demand low overhead and high beamforming gain, theoretically, learning joint-structure and multimodal distribution enables the model to better understand the signal propagation process within complex sites, thereby achieving more accurate beamforming. Nevertheless, the application of GenAI to this problem remains largely unexplored in the existing literature, which raises the following important questions: (1) \textit{How suitable is GenAI in the context of SSBF?} and (2) \textit{to what extent can GenAI improve the beamforming gain?} Therefore, in this paper, we propose a unified design framework for generative SSBF (GenSSBF) and establish a GenSSBF baseline that is fully compatible with the beamforming procedure in current 5G standard. The main contributions are summarized as follows:
\begin{itemize}
	\item We propose a unified GenSSBF framework by considering the fundamental differences between beam generation tasks and image synthesis. The proposed GenSSBF framework consists of site profile, structured wireless prompting module and the generator. The workflow is designed whereby the modules work in concert to generate beams that accurately capture site-specific information while adhering to physical constraints. In contrast to the conventional unstructured discriminative regression based SSBF, GenSSBF jointly learns the internal correlations of the beamforming vector and its interactions with the external propagation environment, thereby significantly enhancing site-specific awareness. 
	\item We propose a \underline{B}eam-\underline{B}rain\underline{S}torm (BBS) solution as an instantiation of the GenSSBF framework. Specifically, the site profile is configured by transforming channel data from spatial domain into a reversible latent space via DFT, thereby reducing the learning difficulty. For the prompting module, the raw wireless prompt is defined as the reference signal received power (RSRP) obtained from probing using a small set of uniformly selected DFT beams. A condition diffusion model is then developed to generate user-specific beams guided by the wireless prompts. The term ``Beam-brainstorm'' reflects the model's ability to generate a set of high-fidelity beams on-demand for different UEs, which can be utilized in the subsequent beam sweeping process. 
	\item Extensive simulation results validate the effectiveness of the proposed BBS solution. Specifically, the BBS outperforms both exhaustive search and discriminative regression-based beam prediction, achieving a higher beamforming gain with significantly lower beam sweeping overhead across various environments (e.g., indoor and outdoor scenarios covering both line-of-sight (LoS)/NLoS link conditions). An important observation is that with limited wireless prompts, brainstorming can significantly boost the achievable beamforming gain, even in low signal-to-noise ratio (SNR) environments. 
\end{itemize}

The rest of this paper is organized as follows. The system model and problem formulation are described in Section \ref{Section2}. In Section \ref{Section3}, the proposed unified design framework for GenSSBF and the BSS solution are introduced. The datasets employed and the simulation results are provided in Section \ref{Section4}, which is followed by our conclusions in Section \ref{Section5}.

\emph{Notations:} Scalars, vectors, and matrices are represented by regular, bold lowercase, and bold uppercase (e.g., x, $\bm{{\rm x}}$ and $\bm{{\rm X}}$) letters, respectively. The set of complex and real numbers are denoted by $\mathbb{C}$ and $\mathbb{R}$, respectively. The transpose and conjugate transpose are denoted by $(\cdot)^T$ and $(\cdot)^H$, respectively. The absolute value and Euclidean norm are denoted by $\vert\cdot\vert$ and $\Vert\cdot\Vert$, respectively. $\mathcal{N}(a,b^2)$ is denoted as a Gaussian distribution with mean $a$ and variance $b^2$. The expectation operator is denoted by $\mathbb{E}[\cdot]$. 

\section{System Model and Problem Formulation}\label{Section2}
Consider a downlink communication system where a BS with an antenna array of $N$ elements serves $K$ single-antenna UEs. A general ray-based channel model with $L$ paths is adopted, which is expressed as
\begin{equation}
	\setlength\abovedisplayskip{3pt}
	\setlength\belowdisplayskip{3pt}
	\bm{{\rm h}}=\sum_{l=1}^L\alpha_l\bm{{\rm a}}(\phi_l^D, \theta_l^D),
\end{equation}
where $\alpha_l$ is the complex gain of the $l$-th path, the azimuth and elevation angles of departure are denoted as $\phi_l^D$ and $\theta_l^D$, respectively, and $\bm{{\rm a}}(\phi_l^D, \theta_l^D)$ represents the array steering vector. For simplicity, we consider a uniform linear array (ULA) in this work where the beam steering is constrained to the azimuth domain\footnote{Our proposed framework can be applied to arrays of arbitrary geometry.}. Accordingly, the steering vector at $\phi_l$ can be written as
\begin{equation}
	\setlength\abovedisplayskip{3pt}
	\setlength\belowdisplayskip{3pt}
	\bm{{\rm a}}(\phi_l) \!=\! \frac{1}{\sqrt{N}}\left[1,{\rm e}^{j\frac{2\pi d}{\lambda}{\rm sin}(\phi_l) }, \cdots, {\rm e}^{j(N_t-1)\frac{2\pi d}{\lambda}{\rm sin}(\phi_l) }\right]^{\rm T},
\end{equation}
with $\lambda$ and $d$ being the carrier wavelength and antenna spacing, respectively.

Given the cost and complexity of fully digital beamforming, especially at high-frequency bands, the BS typically adopts analog-only or hybrid beamforming\cite{heng2022learning}. Accordingly, in this work, it is assumed that BS employs a single radio-frequency (RF) chain, utilizing a network of phase shifters for transmit beamforming. With this architecture, the transmit beamforming vector for a UE can be written as
\begin{equation}\label{eq_beamforming_vector}
	\setlength\abovedisplayskip{3pt}
	\setlength\belowdisplayskip{3pt}
	\bm{{\rm w}}=\frac{1}{\sqrt{N}}[{\rm e}^{j\theta_1}, {\rm e}^{j\theta_2}, \cdots, {\rm e}^{j\theta_{N}}]^{\rm T},
\end{equation}
where $\bm{{\rm w}}\in \mathbb{C}^{N\times 1}$ satisfies the power constraint and each element of $\bm{{\rm w}}$ adheres the constant modulus constraint. Denote $s$ as the transmitted symbol for the UE with the average power constraint $\mathbb{E}[\vert s \vert^2]=1$. Therefore, the received signal at this UE can be given as
\begin{equation}\label{eq_received_signal}
	y(\bm{{\rm w}}) = \sqrt{P_T}\bm{{\rm h}}^{{\rm H}}\bm{{\rm w}}s+n,
\end{equation}
where $P_T$ is the transmit power at BS, and $n\sim \mathcal{CN}(0, \sigma^2)$ is the complex additive noise with noise power $\sigma^2$. The signal-to-noise ratio (SNR) at the UE can be expressed as
\begin{equation}\label{eq_snr}
	\setlength\abovedisplayskip{3pt}
	\setlength\belowdisplayskip{3pt}
	{\rm SNR} = \frac{P_T\vert \bm{{\rm h}}^{{\rm H}}\bm{{\rm w}} \vert^2}{\sigma^2}.
\end{equation}
The ultimate goal is to design a beamforming vector that can maximize the SNR. Nevertheless, due to the complexity and high latency of acquiring instantaneous CSI, the current implementation of 5G adopts a beamforming framework based on beam sweeping, measurements and reporting, where the beams are selected from a predefined codebook. Therefore, for those codebook-based methods, the goal can be converted to find the optimal narrow beam index in the predefined codebook $\bm{{\rm W}}\in \mathbb{C}^{N\times N_W}$ with $N_W$ beams (e.g., DFT codebook or some learnt codebooks), which can be expressed as
\begin{equation}
	\setlength\abovedisplayskip{3pt}
	\setlength\belowdisplayskip{3pt}
	i_{\bm{{\rm W}}}^*\!\!=\!\!\!\!\mathop{{\rm argmax}}\limits_{i=1,2,\cdots,N_W}\left(\frac{P_T\vert \bm{{\rm h}}^{{\rm H}}\bm{{\rm w}}_i \vert^2}{\sigma^2}\right)\!=\!\!\!\mathop{{\rm argmax}}\limits_{i=1,2,\cdots,N_W}(\vert \bm{{\rm h}}^{{\rm H}}\bm{{\rm w}}_i \vert^2).
\end{equation}

However, regardless of being site-agnostic or site-specific, the codebook is common for all UEs within the site. Consequently, the overall achievable beamforming gain is fundamentally bounded by the designed codebook. In this paper, going back to basics, we aim at designing the beamforming vector directly for each individual UE without acquiring full CSI in the deployment stage. Specifically, given limited observations $\bm{{\rm c}}_k$\footnote{The observations can be of different modes, which will be described in Section \ref{Section3.A}.}, the SNR maximization problem is formulated by optimizing a ``beam generator'' $f(\cdot)$ as
\begin{subequations}
	\setlength\abovedisplayskip{3pt}
	\setlength\belowdisplayskip{3pt}
	\begin{align}
		\max_{f} & \quad \vert \bm{{\rm h}}_k^Hf(\bm{{\rm c}}_k)\vert^2 \label{eq_objective_fun}\\
		\mathrm{s.t.} & \quad \Vert f(\bm{{\rm c}}_k) \Vert^2 = 1, \label{eq_constraint1} \\
		& \quad f(\bm{{\rm c}}_k) \in \mathcal{S}, \label{eq_constraint2}
	\end{align}
\end{subequations}
where $\bm{{\rm h}}_k\in \mathbb{C}^{N\times 1}$ is the channel vector of the $k$-UE, constraint (\ref{eq_constraint1}) indicates the normalized power limit for beamforming vector, and $\mathcal{S}$ in (\ref{eq_constraint2}) represents the feasible set constrained by hardware, e.g., the unit modulus constraint on each element of the beamforming vector in (\ref{eq_beamforming_vector}).

\begin{figure*}[t]
	\centering
	\includegraphics[width=6.2in]{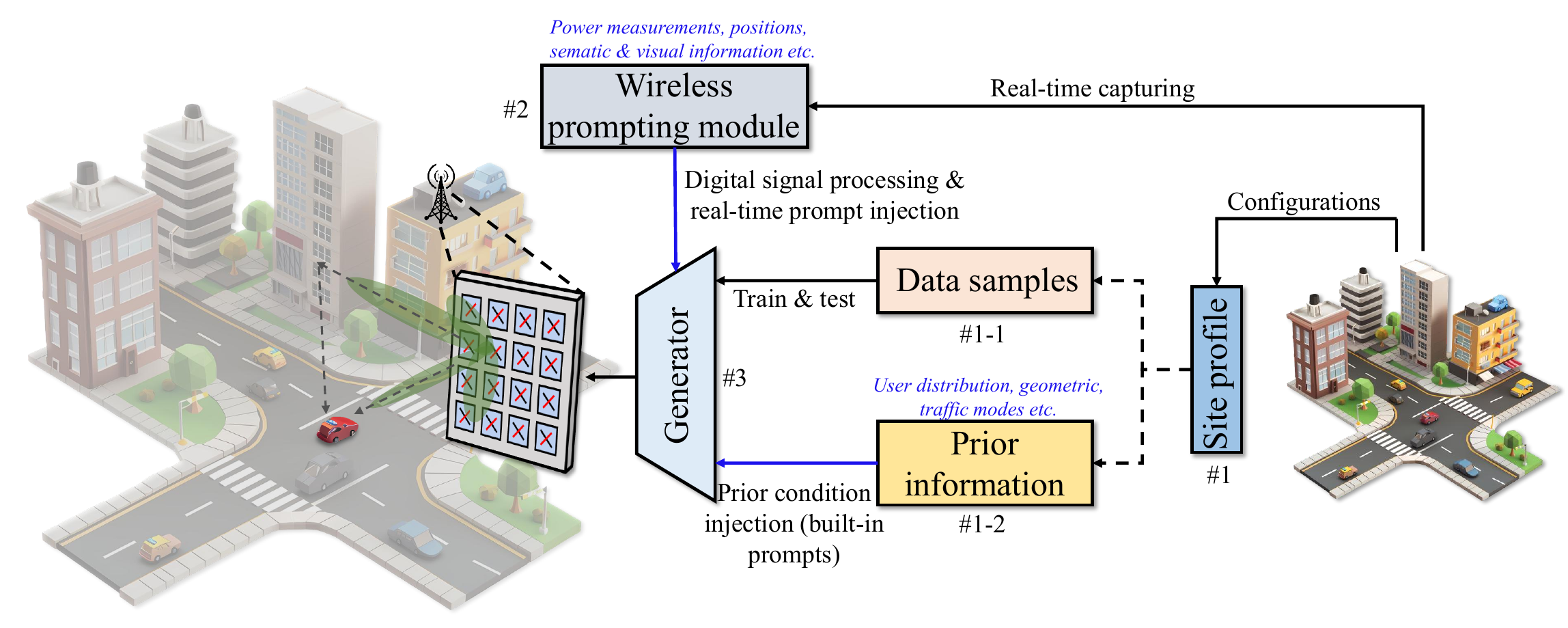}
	\caption{A unified framework for GenSSBF.}
	\label{figure_1}
\end{figure*}

\section{The Proposed Generative Site-Specific Beamforming Design}\label{Section3}
This section details the proposed GenSSBF approach. First, we present in Section \ref{Section3.A} the proposed unified GenSSBF framework. As an instantiation of this framework, the proposed BBS solution is then introduced in Section \ref{Section3.B}.
\subsection{Proposed Unified Framework Design}\label{Section3.A}
Existing SSBF approaches either rely on learnt quantified codebooks or employ discriminative AI model for unstructured beam prediction. The former suffer from gird-mismatch issues, while the beam expressive power of the latter could be limited due to its insufficient understanding of the environment. These critical limitations motivate a paradigm shift from traditional unstructured prediction\cite{torralba2024foundations} to structured generation. Therefore, we propose GenSSBF and a unified framework for it, as shown in Fig. \ref{figure_1}. It can be served as a fundamental and general design pipeline. Typically, the GenSSBF framework consists of three components: 1) Site profile; 2) Wireless prompting module and 3) Generator. 

Specifically, for site profile in the proposed GenSSBF framework (box ``\#1'' in Fig. \ref{figure_1}), it can be configured via historical data or digital twin. A site profile comprises data samples and prior information (e.g., user distribution, geometric knowledge). The data samples are used to train and test the model (box ``\#1-1'' in Fig. \ref{figure_1}), while the prior information can be injected into the generative model as built-in prompts (box ``\#1-2'' in Fig. \ref{figure_1}).

However, prior information alone is not enough for generating user-specific beamforming vectors. Thus, another crucial component is wireless prompting module (box ``\#2'' in Fig. \ref{figure_1}), which is served as a real-time user-specific condition provider. Nevertheless, the textual prompts commonly used in conventional generation tasks fail to convey precise propagation environment related information, while structured textual prompts require extensive expert knowledge\cite{zhou2025large}.  Therefore, GenSSBF requires quantifiable structured wireless prompts to accurately capture site information. To this end, our framework introduces the concept of ``\textit{wireless prompts}''. The wireless prompts can be of different modalities to provide quantifiable site-specific information, such as power measurements from a small probing codebook, position information from global navigation satellite system (GNSS), visual and semantic information (e.g., RGB-D images and 3D point cloud). Furthermore, these multimodal prompts can be integrated into heterogeneous conditions to potentially improve the generation performance.

Having the data samples and prompts, a generative model can be trained offline and then deployed for online inference, as shown in the box labeled ``\#3'' in Fig. \ref{figure_1}. However, GenSSBF requires some unique consideration in designing the generative model. First, in contrast to conventional image generation where pixel value fluctuations are often perceptually tolerable, in beam generation, the phase value of each element in beamforming vector is crucial. It directly determine whether the signals are combined constructively or destructively at the target receiver. Second, the generated beam should be useful in real-world systems, which means that some practical constraints (e.g., phase shifter constraint and the power limit at BS) need to be taken into account during model design. In the next subsection, a detailed solution is described as an instantiation of our proposed GenSSBF framework.

\begin{figure*}[t]
	\centering
	\includegraphics[width=6.2in]{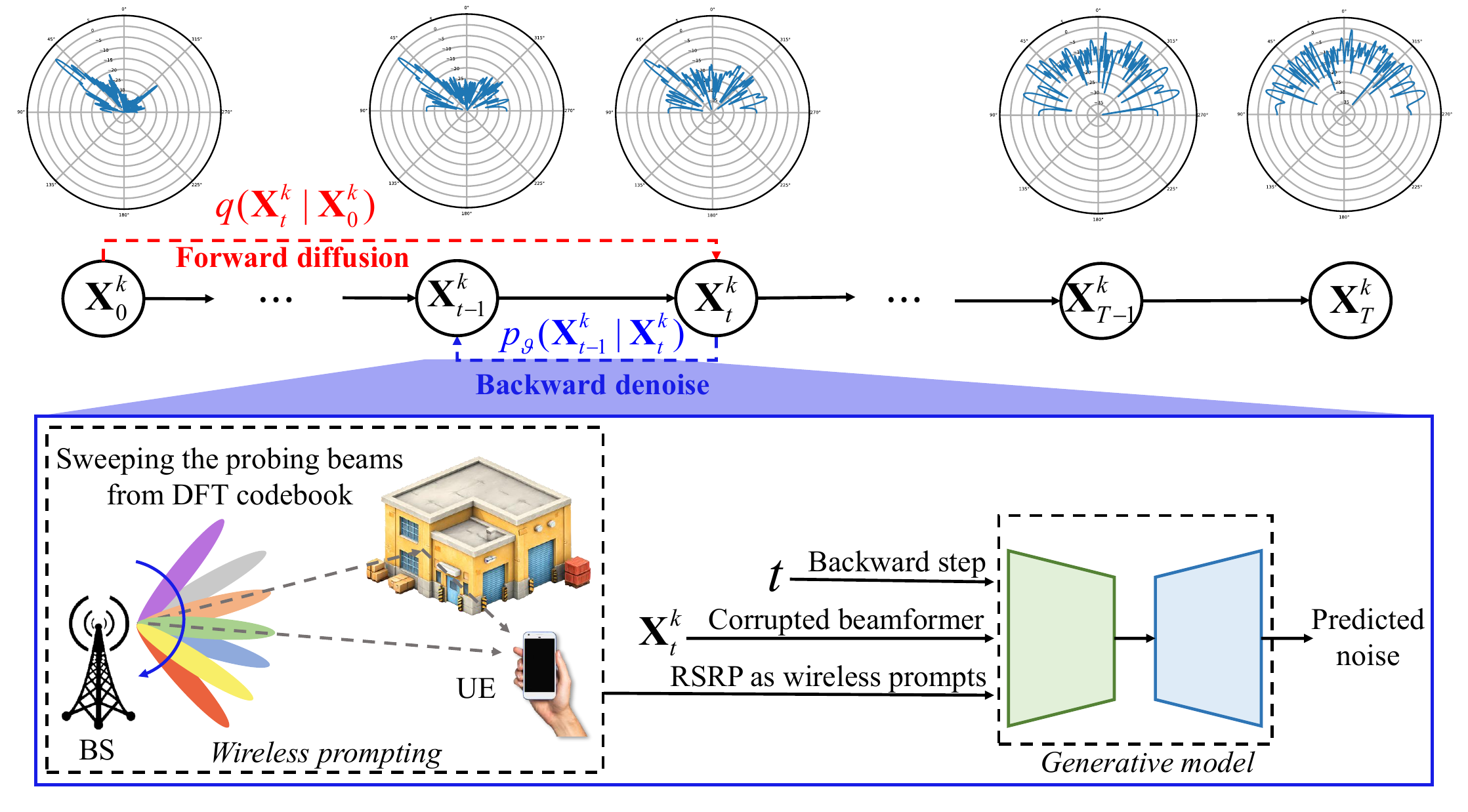}
	\caption{Overall structure of the proposed BBS solution.}
	\label{figure_2}
\end{figure*}

\subsection{BBS for Site-specific Beamforming}\label{Section3.B}
Based on the aforementioned framework, this subsection introduces our proposed beam-brainstorm (BBS) solution for GenSSBF. BBS can implicitly learn the impact of wireless propagation environment on beamforming, representing a paradigm shift from traditional unstructured prediction-based approaches to a generative method. A site-specific and capacity-unbounded beam database for each UE can be represented by the well-trained generative model\footnote{The term ``capacity-unbounded'' means that it parameterizes an infinite space of beamforming vectors for any given UE through the model's weights instead of storing explicitly.}. The term ``brainstorm'' refers to the model's ability to autonomously and on-demand create a compact set of high-fidelity beams, which are then utilized in the subsequent beam sweeping process. Since these generated beams have effectively capture the site-specific information (e.g., the propagation environment), a very small candidate set is adequate, thereby significantly reducing the beam sweeping overhead. The overall framework of the proposed BBS is illustrated in Fig. \ref{figure_2}, in the following, we will detail our BBS solution from three aspects: 1) the formulation of conditional diffusion-based beam generation; 2) the customized model architecture; and 3) the training and deployment algorithms.

\subsubsection{Conditional Diffusion-based Beam Generation}
First, consider the forward diffusion process, which refers to the procedure of gradually corrupting a data sample by adding noise over a series of time steps. In this work, maximum ratio transmission (MRT) beamformer is used as the source of data samples. Specifically, for user $k$ with channel $\bm{{\rm h}}_k\in\mathbb{C}^{N_t\times 1}$, the MRT beamformer is obtained as
\begin{equation}\label{eq_mrt}
	\setlength\abovedisplayskip{3pt}
	\setlength\belowdisplayskip{3pt}
	\bm{{\rm w}}_0^k=\frac{1}{\sqrt{N}}\left[{\rm e}^{j\angle \bm{{\rm h}}_k[1]}, {\rm e}^{j\angle \bm{{\rm h}}_k[2]}, \cdots, {\rm e}^{j\angle \bm{{\rm h}}_k[N]}\right]^{\rm T},
\end{equation}
where $\angle$ represents the phase of a complex number. Instead of directly adding noise to the channel information $\left[\angle \bm{{\rm h}}_k[1], \angle \bm{{\rm h}}_k[2], \cdots, \angle \bm{{\rm h}}_k[N]\right]^{{\rm T}}$ in spatial domain, we leverage the sparse nature of far-field channels in the angular domain to perform forward diffusion process on the discrete Fourier transform (DFT) representation $\bm{{\rm h}}_k^{\mathcal{A}}=\mathcal{F}(\bm{{\rm h}}_k)$ of the spatial channel, with $\mathcal{F}(\cdot)$ being the DFT operator. Therefore, the data sample in this work can be expressed as
\begin{equation}\label{eq_datasample}
	\setlength\abovedisplayskip{3pt}
	\setlength\belowdisplayskip{3pt}
	\bm{{\rm X}}^k_0=
	\begin{bmatrix}
		\angle \bm{{\rm h}}_k^{\mathcal{A}}[1] & \angle \bm{{\rm h}}_k^{\mathcal{A}}[2] & \cdots & \angle \bm{{\rm h}}_k^{\mathcal{A}}[N] \\
		\vert \bm{{\rm h}}_k^{\mathcal{A}}[1] \vert & \vert \bm{{\rm h}}_k^{\mathcal{A}}[2] \vert & \cdots & \vert \bm{{\rm h}}_k^{\mathcal{A}}[N] \vert
	\end{bmatrix}
\end{equation}
where $\vert \cdot \vert$ denotes the amplitude of a complex number \footnote{The amplitude of $\bm{{\rm h}}_k^{\mathcal{A}}$ can be scaled to improve training stability and efficiency.}. Denote $\bm{{\rm X}}_t^k$ as the corrupted, or noisy data for the $k$-th UE at time step $t$. Thus, in a forward diffusion process with a total of $T$ time steps, we will obtain a sequence of corrupted matrices $\{\bm{{\rm X}}_1^k, \bm{{\rm X}}_2^k, \cdots, \bm{{\rm X}}_T^k\}$. Specifically, at each time step $t$, $\bm{{\rm X}}_t^k$ is sampled from the distribution $q(\bm{{\rm X}}_t^k | \bm{{\rm X}}_{t-1}^k)$ which can be expressed as
\begin{equation}\label{eq_pdf_x(t-1)_to_x(t)}
	\setlength\abovedisplayskip{3pt}
	\setlength\belowdisplayskip{3pt}
	q(\bm{{\rm X}}_t^k | \bm{{\rm X}}_{t-1}^k) \!\!=\!\! \prod_{i=1}^2\prod_{j=1}^N\!\mathcal{N}(\bm{{\rm X}}_t^k[i,j];\! \sqrt{1-\beta_t}\bm{{\rm X}}_{t-1}^k[i,j], \beta_t),\!
\end{equation}
where $\beta_t\in (0,1), t=1,2,\cdots,T$, represents the pre-defined noise schedule that controls the rate of noise addition or removal over time. According to the Markov property, the entire forward diffusion process from $\bm{{\rm X}}_0^k$ to $\bm{{\rm X}}_T^k$ is given by
\begin{equation}\label{eq_pdf_x(0)_to_x(T)}
	\setlength\abovedisplayskip{3pt}
	\setlength\belowdisplayskip{3pt}
	q(\bm{{\rm X}}_T^k | \bm{{\rm X}}_{0}^k) = \prod_{t=1}^Tq(\bm{{\rm X}}_t^k | \bm{{\rm X}}_{t-1}^k).
\end{equation}
Eq. (\ref{eq_pdf_x(0)_to_x(T)}) allows us to formulate a recursion, which yields a closed-form relationship from $\bm{{\rm X}}_{0}^k$ to $\bm{{\rm X}}_T^k$ as
\begin{equation}\label{eq_forward_diffusion}
	\setlength\abovedisplayskip{3pt}
	\setlength\belowdisplayskip{3pt}
	\bm{{\rm X}}_T^k = \sqrt{\prod_{t=1}^T(1-\beta_t)}\bm{{\rm X}}_{0}^k+\sqrt{1-\prod_{t=1}^T(1-\beta_t)}\bm{{\rm Z}}_{0\to T},
\end{equation}
where $\bm{{\rm Z}}_{0\to T}$ is the noise matrix of the same shape as $\bm{{\rm X}}_{0}^k$. Each element of $\bm{{\rm Z}}_{0\to T}$ is a Gaussian random variable with zero mean and unit variance. 

\begin{figure}[t]
	\centering
	\includegraphics[width=3.3in]{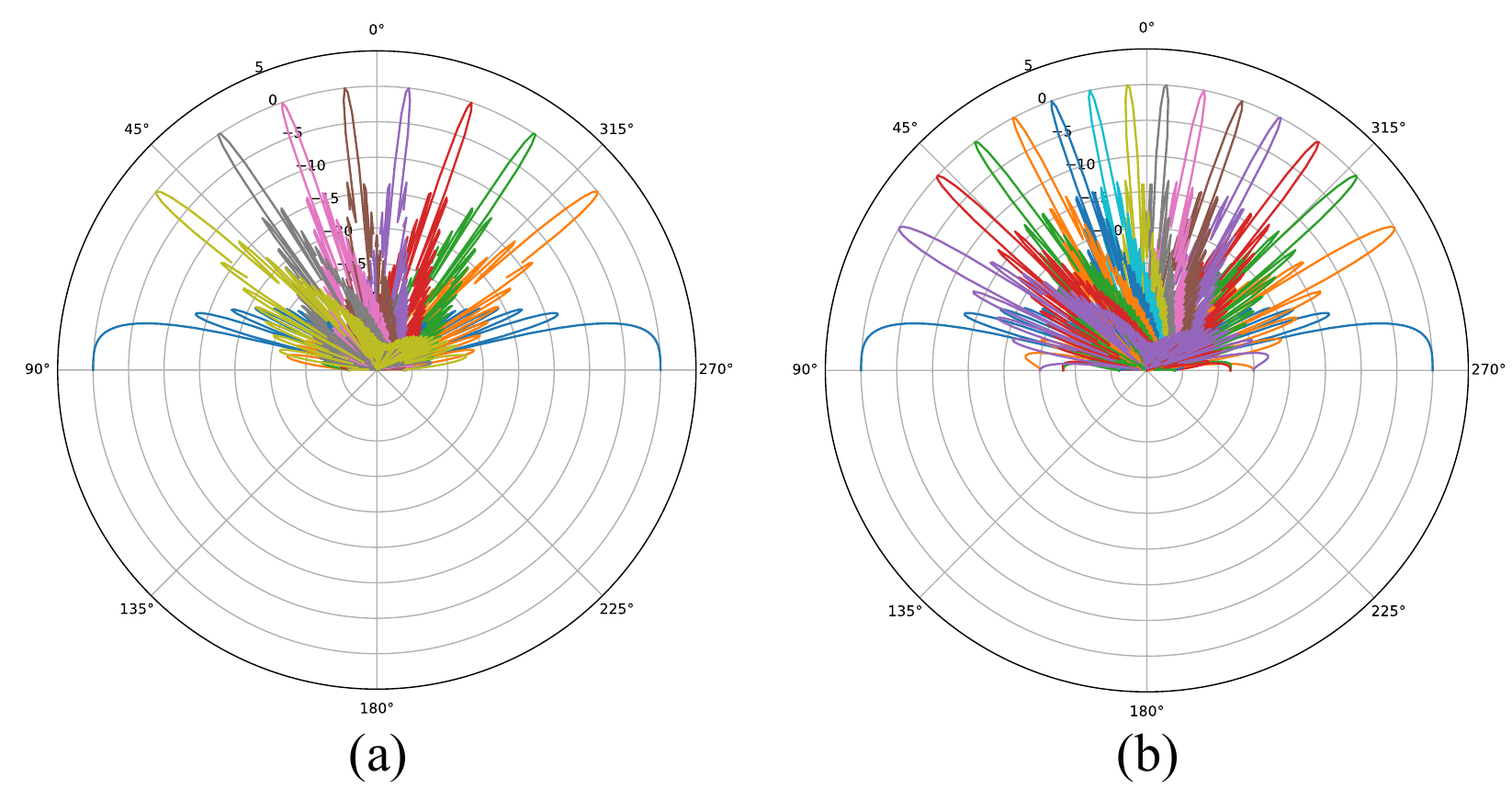}
	\caption{Probing beams for real-time conditioning, (a) $Q$=9 and (b) $Q$=15.}
	\label{figure_3}
\end{figure}

Next, consider the backward denoise process, which refers to the procedure of iteratively denoising $\bm{{\rm X}}_t^k$ to recover the original data sample $\bm{{\rm X}}_{0}^k$. According to the Bayes' theorem, we have
\begin{equation}\label{eq_bayes_theorem}
	\setlength\abovedisplayskip{3pt}
	\setlength\belowdisplayskip{3pt}
	q(\bm{{\rm X}}_{t-1}^k|\bm{{\rm X}}_t^k, \bm{{\rm X}}_0^k)=\frac{q(\bm{{\rm X}}_{t}^k|\bm{{\rm X}}_{t-1}^k, \bm{{\rm X}}_0^k)q(\bm{{\rm X}}_{t-1}^k|\bm{{\rm X}}_0^k)}{q(\bm{{\rm X}}_{t}^k|\bm{{\rm X}}_0^k)}.
\end{equation}
The terms $q(\bm{{\rm X}}_{t}^k|\bm{{\rm X}}_{t-1}^k, \bm{{\rm X}}_0^k)$, $q(\bm{{\rm X}}_{t-1}^k|\bm{{\rm X}}_0^k)$, and $q(\bm{{\rm X}}_{t}^k|\bm{{\rm X}}_0^k)$ in (\ref{eq_bayes_theorem}) can be obtained from (\ref{eq_pdf_x(t-1)_to_x(t)}) and (\ref{eq_pdf_x(0)_to_x(T)}). Consequently, by rearranging, we will have

\begin{equation}\label{eq_pdf_x(t)_to_x(t-1)}
	\setlength\abovedisplayskip{3pt}
	\setlength\belowdisplayskip{3pt}
	q(\bm{{\rm X}}_{t-1}^k|\bm{{\rm X}}_t^k, \bm{{\rm X}}_0^k)=\prod_{i=1}^2\prod_{j=1}^N\mathcal{N}\left(\bm{{\rm X}}_{t-1}^k[i,j];\bm{{\rm U}}_t[i,j], \sigma_t^2\right),
\end{equation}
where $\bm{{\rm U}}_t[i,j]$ can be expressed as
\begin{align}\label{eq_ut[i,j]}
	\bm{{\rm U}}_t[i,j]&=\frac{\sqrt{1-\beta_t}\left(1-\prod_{i=1}^{t-1}(1-\beta_i)\right)}{1-\prod_{i=1}^t(1-\beta_t)}\bm{{\rm X}}_t^k[i,j] \nonumber \\
	& + \frac{\beta_t\sqrt{\prod_{i=1}^{t-1}(1-\beta_i)}}{1-\prod_{i=1}^t(1-\beta_i)}\bm{{\rm X}}_0^k[i,j],
\end{align}
and $\sigma_t^2$ is given by
\begin{equation}\label{eq_sigma^2}
	\setlength\abovedisplayskip{3pt}
	\setlength\belowdisplayskip{3pt}
	\sigma_t^2 = \frac{\beta_t\left(1-\prod_{i=1}^{t-1}(1-\beta_i)\right)}{1-\prod_{i=1}^t(1-\beta_i)}.
\end{equation}
Recall that in (\ref{eq_forward_diffusion}), the relationship between $\bm{{\rm X}}_t^k$ and $\bm{{\rm X}}_0^k$ is
\begin{equation}\label{eq_x_0^k}
	\setlength\abovedisplayskip{3pt}
	\setlength\belowdisplayskip{3pt}
	\bm{{\rm X}}_0^k = \frac{\bm{{\rm X}}_t^k - \sqrt{1-\prod_{i=1}^t(1-\beta_i)}\bm{{\rm Z}}_{0\to t}}{\sqrt{\prod_{i=1}^t(1-\beta_i)}}.
\end{equation}
However, it is noticed that $\bm{{\rm Z}}_{0\to t}$ is only known in the forward diffusion process, thus, a deep neural network $\epsilon_\vartheta$ parameterized by $\vartheta$ is used to predict the noise matrix in the backward denoise process. Furthermore, in order to enable the model to generate meaningful beam for each individual UE, the received power measurements reported from the UE are employed as the real-time wireless prompt to guide the generation process. Specifically, the BS transmits reference signals to the $k$-th UE using $Q$ probing beams that are uniformly selected\footnote{The selection strategy of probing beams can be optimized or even replaced by other modal conditioning mechanisms, which is left for our future work.} from a DFT codebook with $N$ narrow beams, as shown in Fig. \ref{figure_3}. Therefore, based on (\ref{eq_received_signal}), the power measurement vector $\bm{{\rm c}}_k\in \mathbb{R}^{Q\times 1}$ is the condition for generating $\bm{{\rm X}}_{0}^k$ for the $k$-th UE, which can be given by
\begin{equation}\label{eq_RSRP}
	\setlength\abovedisplayskip{3pt}
	\setlength\belowdisplayskip{3pt}
	\bm{{\rm c}}_k=\left[\vert y_k(\bm{{\rm w}}^{{\rm DFT}}_1)\vert^2, \vert y_k(\bm{{\rm w}}^{{\rm DFT}}_2)\vert^2, \cdots, \vert y_k(\bm{{\rm w}}^{{\rm DFT}}_Q)\vert^2\right]^{{\rm T}},
\end{equation}
where $\bm{{\rm w}}^{{\rm DFT}}_q$ represents the beamforming vector of the $q$-th narrow beam selected from the DFT codebook. Thus, the predicted noise matrix at time step $t$ can be expressed as

\begin{equation}\label{eq_nn_noise_predictor}
	\setlength\abovedisplayskip{3pt}
	\setlength\belowdisplayskip{3pt}
	\epsilon_\vartheta(\bm{{\rm X}}_t^k, \bm{{\rm c}}_k, t) \approx \bm{{\rm Z}}_{0\to t}.
\end{equation}
By substituting (\ref{eq_nn_noise_predictor}) and (\ref{eq_x_0^k}) into (\ref{eq_ut[i,j]}) yields
\begin{equation}\label{eq_ut'}
	\setlength\abovedisplayskip{3pt}
	\setlength\belowdisplayskip{3pt}
	\bm{{\rm U}}_t'= \frac{1}{\sqrt{1-\beta_t}}\left(\bm{{\rm X}}_{t}^k-\frac{\beta_t}{\sqrt{1-\prod_{i=1}^t(1-\beta_i)}}\epsilon_\vartheta(\bm{{\rm X}}_t^k, \bm{{\rm c}}_k, t)\right).
\end{equation}
Therefore, the conditional distribution $q(\bm{{\rm X}}_{t-1}^k|\bm{{\rm X}}_t^k, \bm{{\rm X}}_0^k)$ in eq. (\ref{eq_pdf_x(t)_to_x(t-1)}) can be approximated using a parameterized model $p_\vartheta(\bm{{\rm X}}_{t-1}^k|\bm{{\rm X}}_t^k)$, which can be expressed as
\begin{equation}\label{eq_singlestep_denoise}
	\setlength\abovedisplayskip{3pt}
	\setlength\belowdisplayskip{3pt}
	p_\vartheta(\bm{{\rm X}}_{t-1}^k|\bm{{\rm X}}_t^k)=\prod_{i=1}^2\prod_{j=1}^N\mathcal{N}\left(\bm{{\rm X}}_{t-1}^k;\bm{{\rm U}}_t'[i,j], \sigma_t^2\right).
\end{equation}
The generative distribution $p_\vartheta(\bm{{\rm X}}_{0}^k)$ for the $k$-th UE is defined through the reverse Markov chain from $\bm{{\rm X}}_{T}^k$ to $\bm{{\rm X}}_{0}^k$:
\begin{equation}\label{eq_pdf_X0}
	\setlength\abovedisplayskip{3pt}
	\setlength\belowdisplayskip{3pt}
	p_\vartheta(\bm{{\rm X}}_{0}^k) = q(\bm{{\rm X}}_{T}^k)\prod_{t=1}^Tp_\vartheta(\bm{{\rm X}}_{t-1}^k|\bm{{\rm X}}_t^k),
\end{equation}
where at each time step $t=1,2,\cdots,T$, the denoising update is performed using deterministic approximation as $\bm{{\rm X}}_{t-1}^k = \bm{{\rm U}}_t'$. This is because beamforming requires more precise generation of the phase and amplitude of the complex number compared to traditional image generation tasks, which are less sensitive to pixel-level fluctuations. Nevertheless, in BBS, the diversity of beam generation can still be preserved, i.e., different samples of $\bm{{\rm X}}_{T}^k$ will correspond to different beam patterns for the $k$-th UE. 

Consequently, after repeating the denoise step in (\ref{eq_singlestep_denoise}) from $t=T$ to $t=1$, $\bm{{\rm X}}_{0}^k$ can be reconstructed, and the $n$-th element of the beamforming vector for user $k$ can be expressed as
\begin{equation}\label{eq_idft}
	\setlength\abovedisplayskip{3pt}
	\setlength\belowdisplayskip{3pt}
	\bm{{\rm w}}_0^k[n]=\frac{1}{\sqrt{N}}{\rm e}^{j\angle \mathcal{F}^{-1}\left(\bm{{\rm X}}_{0}^k[2,n]{\rm e}^{j\bm{{\rm X}}_{0}^k[1,n]}\right)},
\end{equation}
where $\mathcal{F}^{-1}(\cdot)$ is the IDFT operator.

\subsubsection{Model Architecture}
\begin{figure}[t]
	\centering
	\includegraphics[width=3.4in]{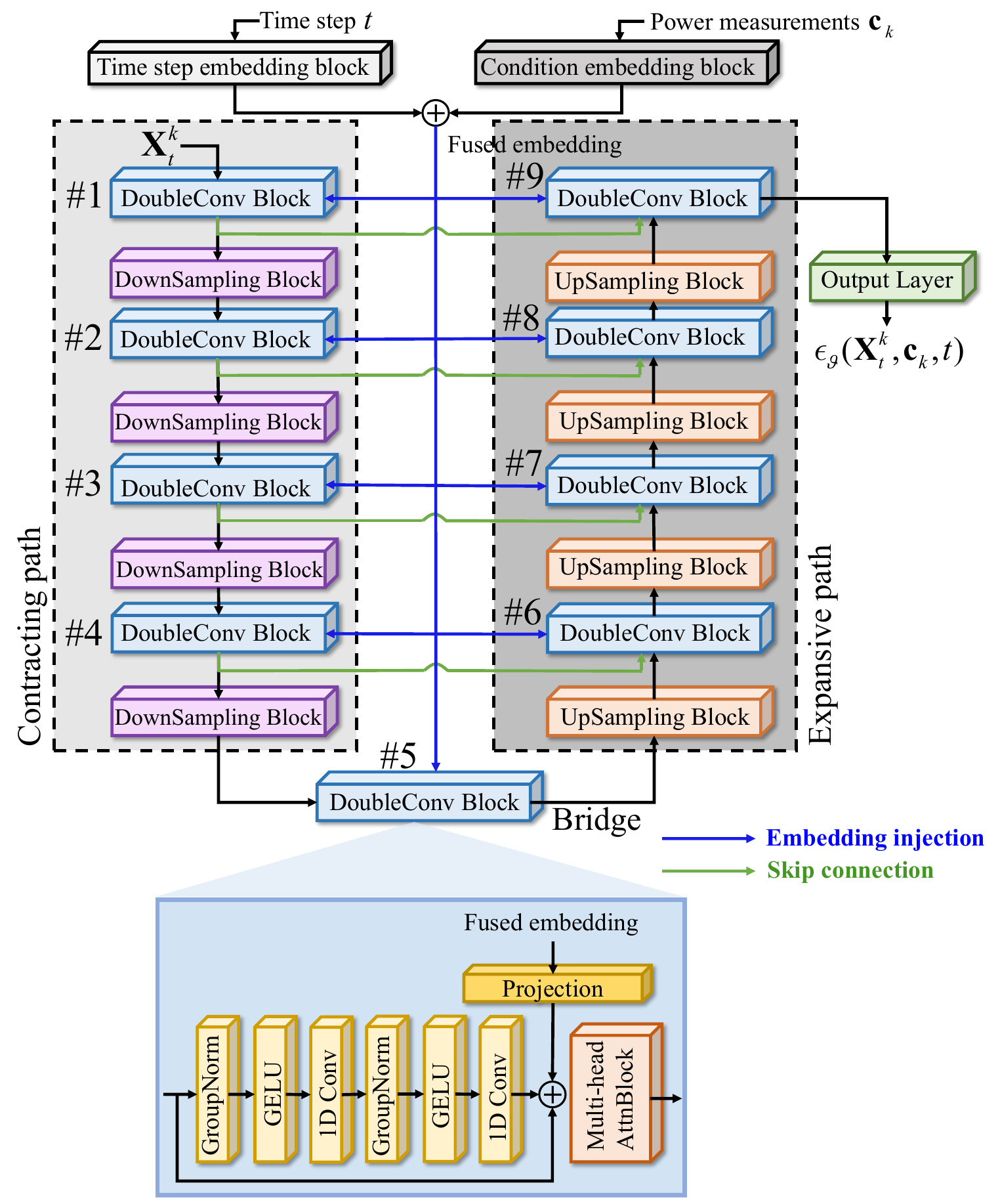}
	\caption{The model architecture.}
	\label{figure_4}
\end{figure}

A modified U-Net model is employed in BBS to denoise the beamformer. Specifically, as illustrated in Fig. \ref{figure_4}, the model consists of three main building blocks: ``DoubleConv Block'', ``DownSampling Block'' and ``UpSampling Block'', which collectively define the contracting path, bridge, and expanding path of the model. 

In ``DoubleConv Block'' (as shown in the blue box of Fig. \ref{figure_4}), two 1D convolution layers (kernel size = 1, padding = 1) are cascaded, each preceded by a Gaussian error linear unit (GELU) activation layer\cite{hendrycks2016gaussian}. The output of the second 1D convolution layer is summed--via a residual connection--with a projected fused embedding and the original input of this ``DoubleConv Block''. Then, this result is fed into a multi-head self-attention block to produce a beam feature map. In the contracting path, every ``DownSampling Block'' employs 1D max pooling to reduce the length of the beam feature map by half. Conversely, in the expanding path, each ``UpSampling Block'' adopts 1D linear interpolation to increase the length. The upsampled beam feature map from the ``UpSampling Block'' is concatenated with the corresponding feature map from the contracting path at the same stage, as indicated by the green arrow in Fig. \ref{figure_4}. The combined feature map is then served as the input to the next ``DoubleConv Block'' in the expanding path. The denoising process is guided by the time step $t$ and the power measurement vector $\bm{{\rm c}}_k$. A sinusoidal positional encoding is used to obtain the time step embedding, while condition embedding is derived through a two-layer MLP with GELU activation function. Finally, the output layer--implemented as a 1D convolution layer--predicts the noise as defined in (\ref{eq_nn_noise_predictor}).

\subsubsection{Training and Deployment}
In the offline training stage, the dataset is $\mathcal{D}=(\bm{{\rm h}}_d, \bm{{\rm X}}_0^d, \bm{{\rm c}}_d)_{d=1}^D$, where $D$ is the total number of activated points of the site whose channel vectors are obtained through ray-tracing simulators. It should be noted that since no prior information about the noise is available in offline training phase, the vector $\bm{{\rm c}}_d$ is constructed from noiseless power measurements with its element given by $\vert \bm{{\rm h}}_d^{{\rm H}}\bm{{\rm w}}_q^{{\rm DFT}}\vert^2$. At each training epoch, the whole dataset is split into mini-batches with the batch size being $B (1\leq B \leq D)$. For each data sample $\bm{{\rm X}}_0^b$ in a batch, we randomly generate an integer time step $t_b$ from $(0, T]$, and corrupt the original data sample $\bm{{\rm X}}_0^b$ by applying $t_b$ steps of the forward diffusion process according to (\ref{eq_forward_diffusion}) to obtain $\bm{{\rm X}}_{t_b}^b$. Denote this corrupted batch as $\bm{{\rm X}}^b_t\in \mathbb{R}^{B\times 2\times N}$, then feed $\bm{{\rm X}}^b_t$, a batch of time steps $\bm{{\rm t}}_B \in \mathbb{Z}_+^{B\times 1}$, and a batch of conditions $\bm{{\rm c}}_B\in \mathbb{R}^{B\times Q}$ into the model to get the prediction $\epsilon_\vartheta(\bm{{\rm X}}_{t}^b, \bm{{\rm t}}_B, \bm{{\rm c}}_B)$ of the injected noise. The mean-squared error (MSE) loss of noise prediction is employed for updating the parameters of the model, which can be expressed as
\begin{equation}\label{eq_loss}
	\setlength\abovedisplayskip{3pt}
	\setlength\belowdisplayskip{3pt}
	\mathcal{L} = \mathbb{E}\left[\Vert \bm{{\rm Z}}_B - \epsilon_\vartheta(\bm{{\rm X}}_{t}^b, \bm{{\rm t}}_B, \bm{{\rm c}}_B) \Vert^2\right],
\end{equation}
where $\bm{{\rm Z}}_B\in\mathbb{R}^{B\times 2\times N}$ is the injected noise matrix in forward diffusion process. Therefore, the parameters $\vartheta$ of the denoise NN can be updated via gradient descent as $\vartheta \leftarrow \vartheta - l_r \nabla_\vartheta \mathcal{L}$, where $l_r$ is the learning rate.

\begin{algorithm}[t]
	\caption{Offline training in BBS}
	\label{algorithm1}
	\begin{algorithmic}
		\STATE \textbf{Initialize}: number of epochs ``MAX\_EPOCHS'', number of probing beams $Q$, learning rate $l_r$, noise schedule $\{\beta_t\}_{t=1}^T$, batch size $B$, diffusion steps $T$, the parameters $\vartheta$ of the modified U-Net.
		\STATE \textbf{/* \textit{Site profile configuration} */}:
		\FOR{each activated UE $d$ in dataset}
		\STATE Obtain the channel information in both spatial domain $\bm{{\rm h}}_d$ and latent space $\bm{{\rm h}}_d^{\mathcal{A}} = \mathcal{F}(\bm{{\rm h}}_d)$
		\STATE Obtain $\bm{{\rm X}}_0^d$ based on (\ref{eq_datasample})
		\STATE Obtain noiseless RSRP vector $\bm{{\rm c}}_d$
		\ENDFOR
		\STATE Construct the training set $\mathcal{D}=(\bm{{\rm h}}_d, \bm{{\rm X}}_0^d, \bm{{\rm c}}_d)_{d=1}^D$
		\STATE \textbf{/* \textit{Offline training} */}:
		\FOR{epoch=1 to MAX\_EPOCHS}
		\FOR{batch $(\bm{{\rm h}}_b, \bm{{\rm X}}_0^b, \bm{{\rm c}}_b)_{b=1}^B\sim \mathcal{D}$}
		\STATE Sample a batch of time steps $\bm{{\rm t}}_B \in \mathbb{Z}_+^{B\times 1}$ from $(0, T]$
		\STATE Corrupt the batch data using $\bm{{\rm t}}_B$ based on (\ref{eq_forward_diffusion})
		\STATE Conduct noise prediction based on (\ref{eq_nn_noise_predictor})
		\STATE Compute the loss based on (\ref{eq_loss})
		\STATE Update $\vartheta$ via gradient descent
		\ENDFOR
		\ENDFOR
	\end{algorithmic}
\end{algorithm}

\begin{algorithm}[t]
	\caption{Online inference for $k$-th UE}
	\label{algorithm2}
	\begin{algorithmic}
		\STATE \textbf{Initialize}: brainstorm number $M$
		\STATE \textbf{/* \textit{Wireless prompting} */}:
		\STATE Uniformly select $Q$ narrow beams in DFT codebook
		\STATE Conduct beam sweeping and get RSRP vector $\bm{{\rm c}}_k$
		\STATE Initialize $M$ Gaussian noise matrices $\{\bm{{\rm X}}_T^{k,m}\}_{m=1}^{M}$ based on (\ref{eq_initial_noise_matrix})
		\STATE \textbf{/* \textit{Brainstorm} */}:
		\FOR{$m=1$ to $M$}
		\FOR{$t=T$ to $1$}
		\STATE Obtain $\bm{{\rm X}}_{t-1}^{k,m}$ by backward denoise based on (\ref{eq_ut'})
		\ENDFOR
		\ENDFOR
		\STATE Conduct beam sweeping using $\{\bm{{\rm X}}_0^{k,m}\}_{m=1}^{M}$ and get RSRP
		\STATE Obtain the optimal beamforming vector $\bm{{\rm X}}_0^{k,*}$
	\end{algorithmic}
\end{algorithm}

\begin{figure}[t]
	\centering
	\includegraphics[width=3.2in]{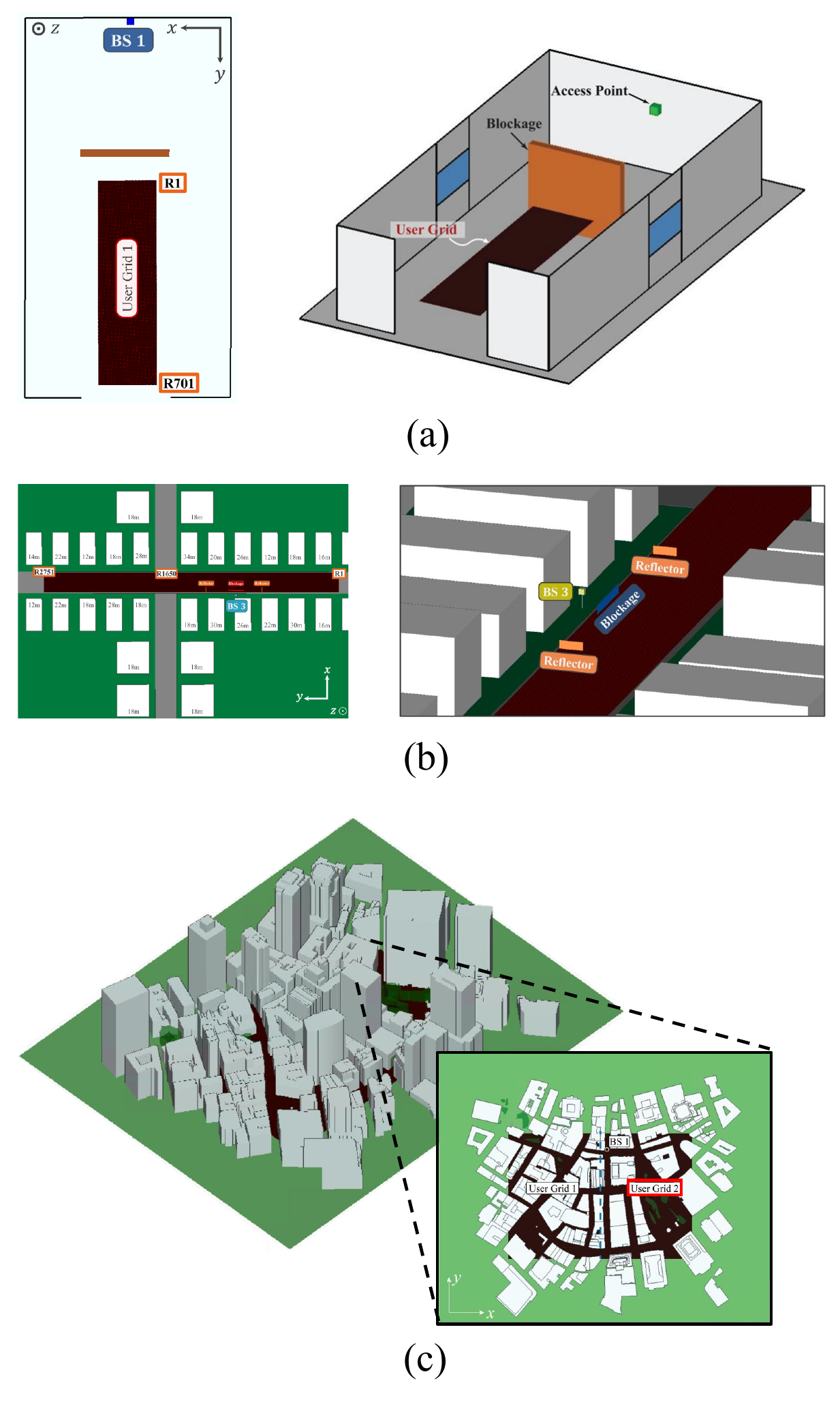}
	\caption{Illustration of the ray-tracing environments, (a) I2\_28B, (b) O1B\_28 and (c) Boston5G\_28.}
	\label{figure_5}
\end{figure}

After training, for a certain UE $k$, we initialize $M$ Gaussian noise matrices $\{\bm{{\rm X}}_T^{k,m}\}_{m=1}^{M}$ where 
\begin{equation}\label{eq_initial_noise_matrix}
	\setlength\abovedisplayskip{3pt}
	\setlength\belowdisplayskip{3pt}
	q(\bm{{\rm X}}_T^{k,m}) = \prod_{i=1}^2\prod_{j=1}^N\mathcal{N}(\bm{{\rm X}}_T^{k,m}[i,j]; 0, 1).
\end{equation}
For each $\bm{{\rm X}}_T^{k,m}$, an iterative backward denoise process is carried out based on (\ref{eq_ut'}) to obtain $\bm{{\rm X}}_0^{k,m}$. Subsequently, a lossless inverse mapping based on IDFT is applied to reconstruct $\bm{{\rm X}}_0^{k,m}$ into a beamforming vector $\bm{{\rm w}}_0^{k,m}$, as formulated in (\ref{eq_idft}). After sequentially applying $T$ denoising steps to $M$ different Gaussian noise matrices, $M$ beamforming vectors customized for the $k$-th UE are generated. The BS then sweeps through these $M$ beams and, based on the feedback from the UE, selects the beam with the highest received power.

Our proposed framework eliminates the need for a pre-designed global codebook. By leveraging the generative capability of the conditional diffusion model, it enables implicitly envisioning an unlimited number of site-specific beams for each individual UE. During deployment, the model only needs to ``brainstorm'' $M$ beams, where $M$ can typically be very small, thereby keeping the beam sweeping overhead $O$ at $Q+M$ (when $M>1$)\footnote{When $M=1$, the beam sweeping overhead is $Q$.}. The offline training and online inference of the proposed BBS are summarized in \textbf{Algorithm} \ref{algorithm1} and \textbf{Algorithm} \ref{algorithm2}, respectively. In section \ref{Section5}, we demonstrate that the proposed method achieves significantly higher beamforming gain while drastically reducing the overhead.

\begin{remark}
	\label{remark1}
	The proposed BBS solution essentially follows the ``sweeping-measuring-reporting'' BA procedure adopted in current 5G standard. Since the probing beams for obtaining the wireless prompts are uniformly selected form the DFT-codebook, they can be transmitted using ``always-on'' SSBs. The generated user-specific beams can be transmitted using CSI-RSs. Consequently, the proposed BBS solution can be seamless integration into current 5G systems, requiring no changes to the standardized protocols.
\end{remark}

\begin{table}[t]
	\caption{Simulation Parameters}
	\label{table1}
	\centering
	\begin{tabular}{|c|c|}
		\hline
		Name of scenario & I2\_28B, O1B\_28 and Boston5G\_28\\
		BS antenna & 64$\times$1 ULA \\
		Antenna spacing & Half-wavelength spacing \\
		UE antenna & Single \\
		Antenna element & Isotropic \\
		Carrier frequency & 28 GHz \\
		Number of paths & 5 \\
		Total number of data samples & 100k \\
		Training set ratio & 80\% \\
		Learning rate & $10^{-4}$ \\
		Batch size $B$ & 32 \\
		Diffusion steps $T$ & 1000 \\
		Number of epochs & 300 \\
		EMA decay coefficient & 0.995 \\
		\hline
	\end{tabular}
\end{table}

\section{Dataset and Simulation Results}\label{Section4}
\subsection{Dataset}\label{Section4.A}
Accurate datasets are helpful to analyze how the proposed algorithm performs in practical environments. Three different scenarios from the public DeepMIMO dataset\cite{alkhateeb2019generic} are adopted to comprehensively consider various propagation environments, such as indoors and outdoors, LoS and NLoS links, etc. The channel data in DeepMIMO dataset are generated via ray-tracing using a state-of-the-art commercial-grade software. Specifically, in the simulation, we use DeepMIMO I2\_28B, O1B\_28, and Boston5G\_28 to evaluate the performance of the BBS solution, which are described in the following.

\subsubsection{DeepMIMO I2\_28B Scenario}
The DeepMIMO I2\_28B dataset\footnote{ \href{https://www.deepmimo.net/scenarios/v4/i2_28b}{https://www.deepmimo.net/scenarios/v4/i2\_28b}} models an indoor scenario where none of the users has a LOS connection with the BS, as illustrated in Fig. \ref{figure_5}(a). The BS is placed on the inside wall of the room, and a total of 140901 users are distributed in the brown-colored area located behind the blockage. The carrier frequency is 28 GHz.

\subsubsection{DeepMIMO O1B\_28 Scenario}
The DeepMIMO O1B\_28 dataset\footnote{ \href{https://www.deepmimo.net/scenarios/v4/o1b_28}{https://www.deepmimo.net/scenarios/v4/o1b\_28}} models an outdoor street environment with blockage and reflections, as illustrated in Fig. \ref{figure_5}(b). Specifically, a portion of the O1\_28 dataset corresponding to BS \#3 and user grid \#1 is selected. Therefore, the total population of 484264 users consists of those with LoS links to the BS and those with NLoS links. The carrier frequency is 28 GHz.

\subsubsection{DeepMIMO Boston5G\_28 Scenario}
The DeepMIMO Boston5G\_28 dataset\footnote{ \href{https://www.deepmimo.net/scenarios/v4/boston5g_28}{https://www.deepmimo.net/scenarios/v4/boston5g\_28}} features a mix of high-rise buildings, streets and open spaces in downtown Boston, USA. Specifically, the BS is located at a height of 15 meters, and the user grid \#2 with 102762 activate users is selected as shown in Fig. \ref{figure_5}(c). The carrier frequency is 28 GHz.

\begin{figure*}[t]
	\centering
	\includegraphics[width=7.0in]{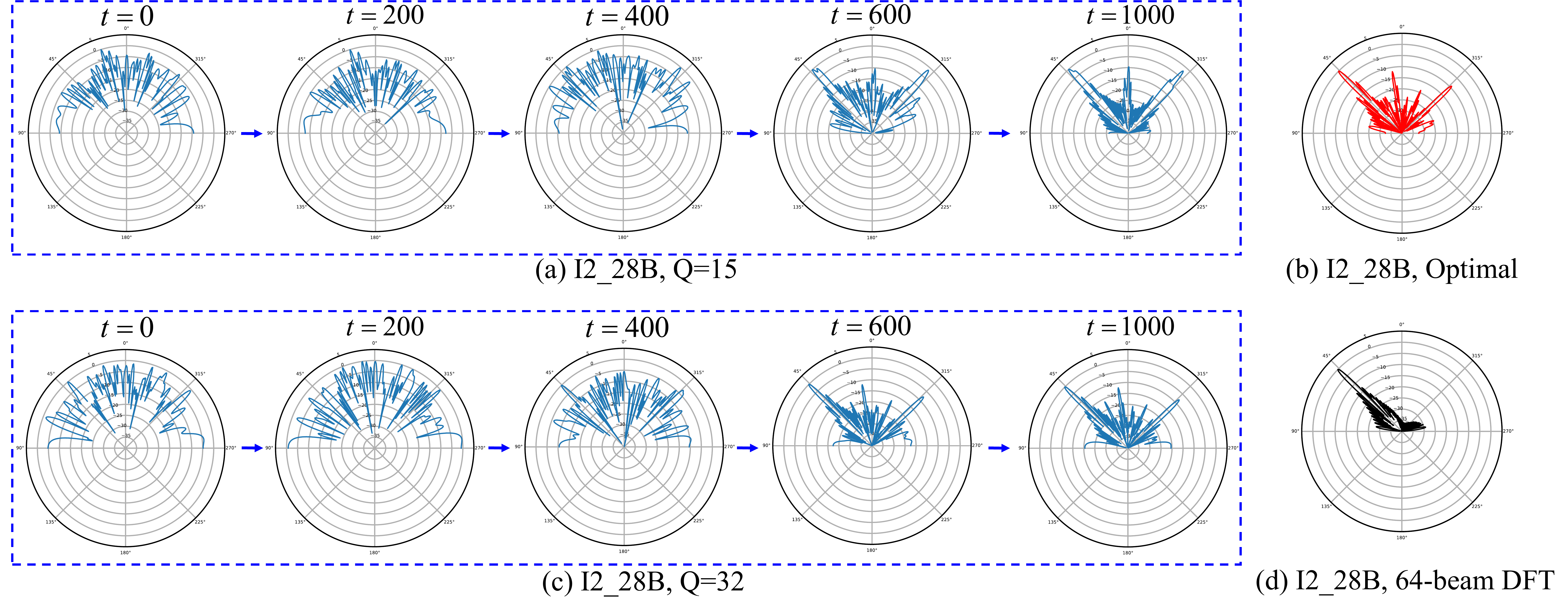}
	\caption{Comparison of the beam pattern in I2\_28B scenario, obtained via (a) BSS with $Q=15$, (b) MRT, (c) BBS with $Q=32$, and (d) exhaustively searching the 64-beam DFT codebook}
	\label{figure_6}
\end{figure*}

\begin{figure*}[t]
	\centering
	\includegraphics[width=7.0in]{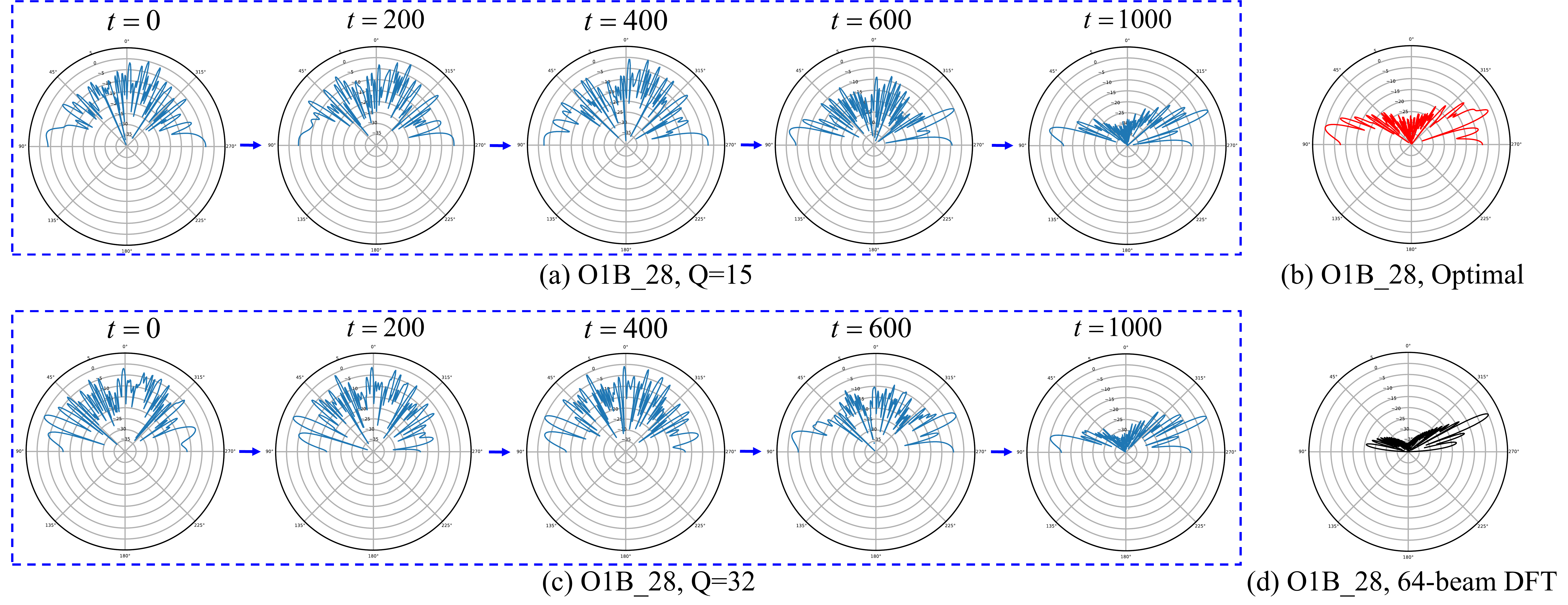}
	\caption{Comparison of the beam pattern in O1B\_28 scenario, obtained via (a) BSS with $Q=15$, (b) MRT, (c) BBS with $Q=32$, and (d) exhaustively searching the 64-beam DFT codebook}
	\label{figure_7}
\end{figure*}

\subsection{Simulation Results}
In this section, we evaluate the performance of the proposed BBS solution under the scenarios described in Section \ref{Section4.A} For each scenario, 80\% of the total data samples of 100k are used for training, and the remaining 20\% for testing. The total diffusion time steps $T$ is set to be 1000. The condition embedding module is a two-layer MLP with 256 neurons in both its hidden and output layers, using a GELU activation function between them. The time step $t$ is also projected into a 256-dimensional embedding space using sinusoidal positional encoding. For denoising NN, in the contracting path and bridge, each DoubleConv block (DoubleConv Block \#1-\#5 in Fig. \ref{figure_4}) processes input with channel dimensions of $[1,64,128,256,512]$ and outputs features with channel dimensions of $[64,128,256,512,1024]$, respectively. Correspondingly, in the expansive path (DoubleConv Block \#6-\#9), the channel dimensions of the output are $[512, 256, 128, 64]$. The number of heads in the multi-head attention block is 4. The model is trained for 300 epochs using the Adam optimizer with the learning rate and batch size being $10^{-4}$ and 32, respectively. Additionally, exponential moving average (EMA) is adopted to update the parameters of the model with decay coefficient of 0.995. The detail parameters are summarized in Table \ref{table1}.

\begin{figure*}[t]
	\centering
	\includegraphics[width=7.0in]{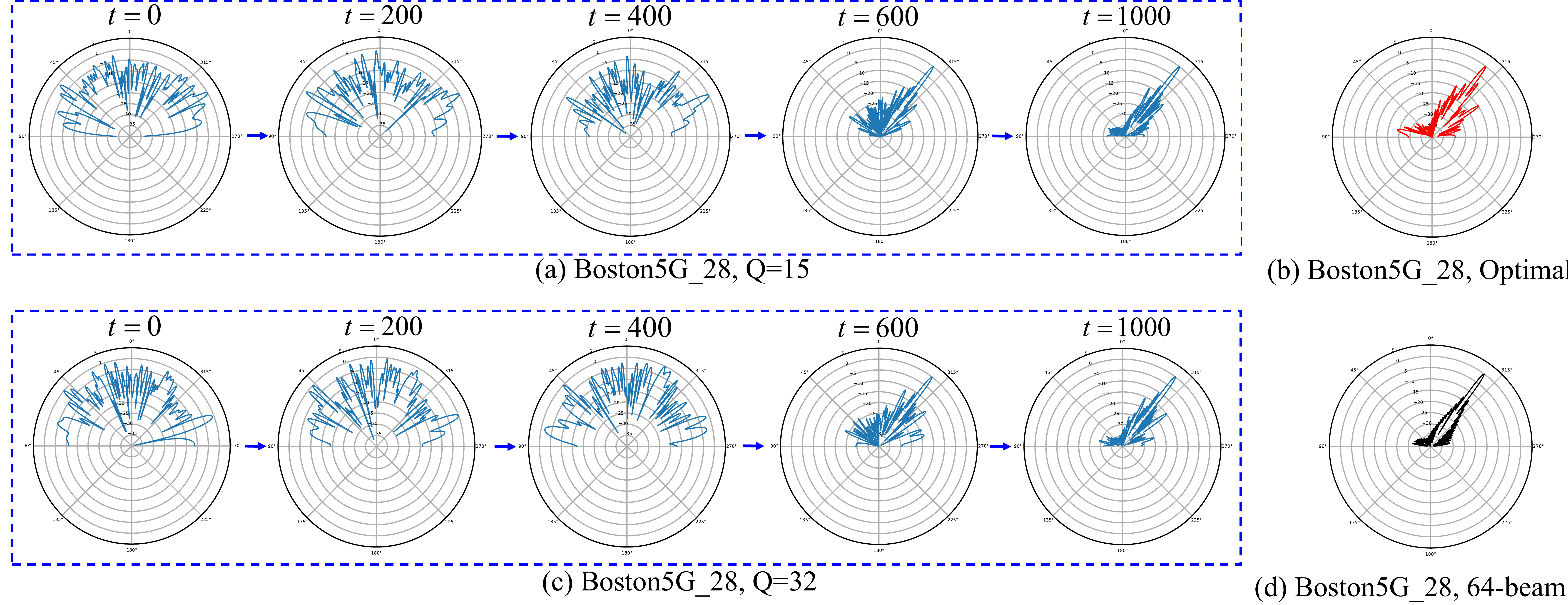}
	\caption{Comparison of the beam pattern in Boston5G\_28 scenario, obtained via (a) BSS with $Q=15$, (b) MRT, (c) BBS with $Q=32$, and (d) exhaustively searching the 64-beam DFT codebook}
	\label{figure_8}
\end{figure*}

\subsection{Analysis of the Generated Beam Pattern}
Before performing a quantitative analysis, we first conduct an intuitive examination of the generated beam pattern to extract potential design insights. Fig. \ref{figure_6}, \ref{figure_7} and \ref{figure_8} illustrate the generation processes, optimal beam patterns, the DFT-beam patterns for I2\_28B, O1B\_28 and Boston5G\_28 scenarios, respectively. As depicted in the dashed blue box in Fig. \ref{figure_6}, \ref{figure_7} and \ref{figure_8}, the beam generation process is shown with denoise steps\footnote{It should be noted that here, the index $t$ refers to the number of backward denoising steps, where $t=400$ corresponds to 400 steps of the denoising process}. It can be observed that the beamforming vector evolves from an initially random pattern to one with clear directional characteristics as the denoising steps progress. Notably, despite being initialized with random matrices sampled from the same Gaussian distribution, the proposed BBS generates distinctly different beam patterns under different wireless prompts across the three scenarios. This result demonstrates that the model has strong capability to capture site-specific information and good contextual awareness of the deployment environment.

\begin{figure}[t]
	\centering
	\includegraphics[width=3.5in]{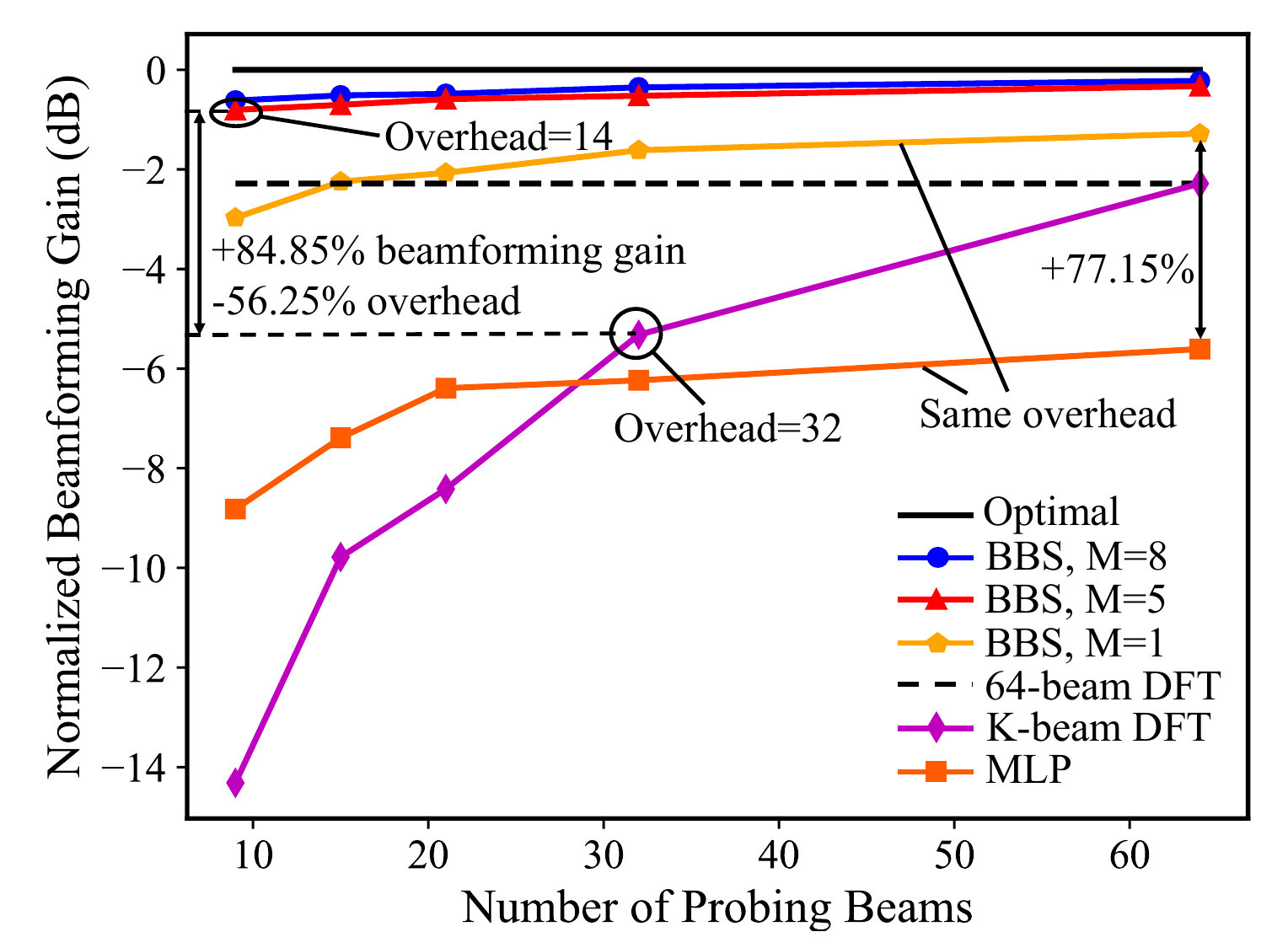}
	\caption{Normalized beamforming gain vs. $Q$ in I2\_28B scenario.}
	\label{figure_9}
\end{figure}

\begin{figure}[t]
	\centering
	\includegraphics[width=3.5in]{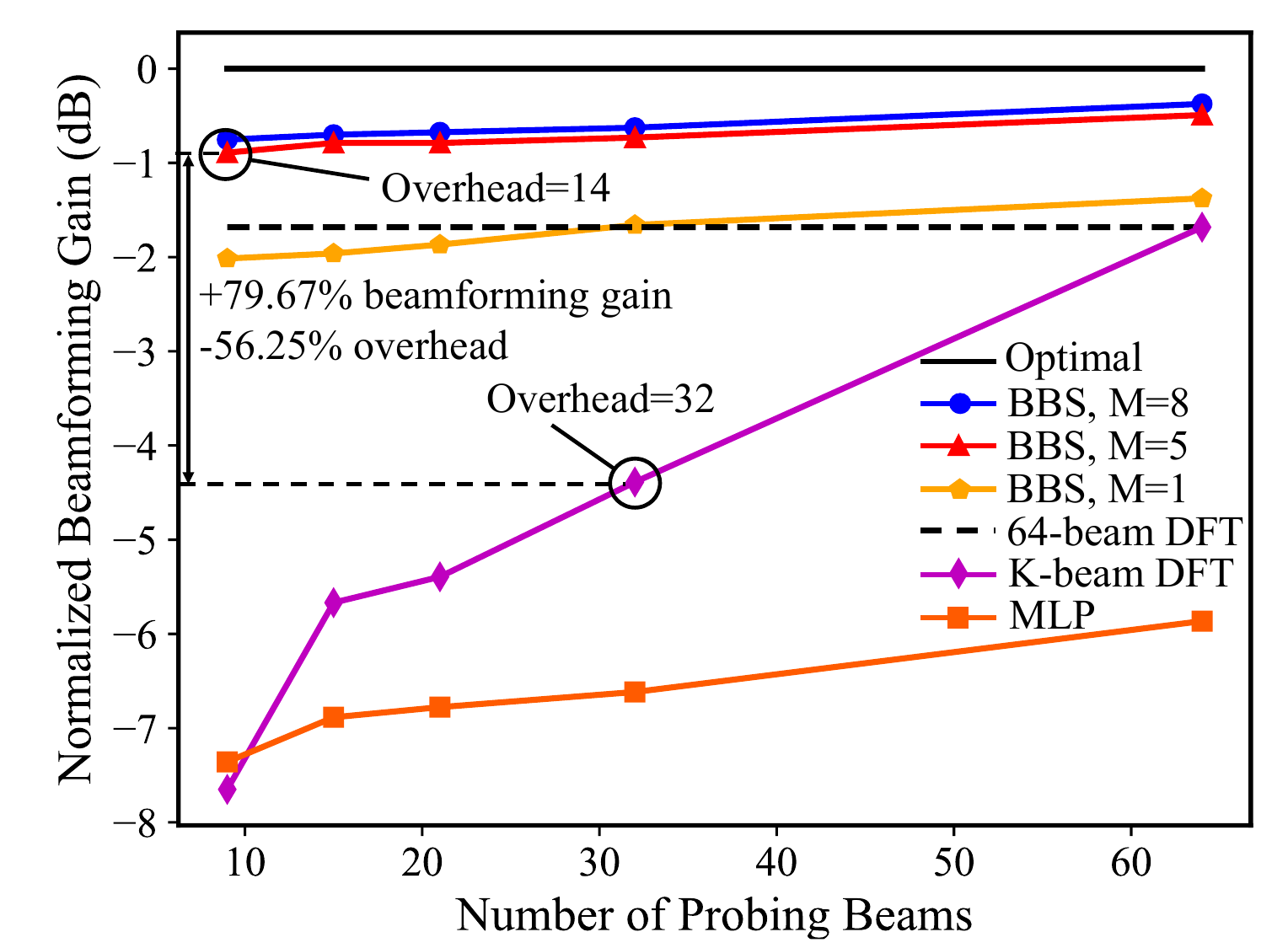}
	\caption{Normalized beamforming gain vs. $Q$ in O1B\_28 scenario.}
	\label{figure_10}
\end{figure}

\begin{figure}[t]
	\centering
	\includegraphics[width=3.5in]{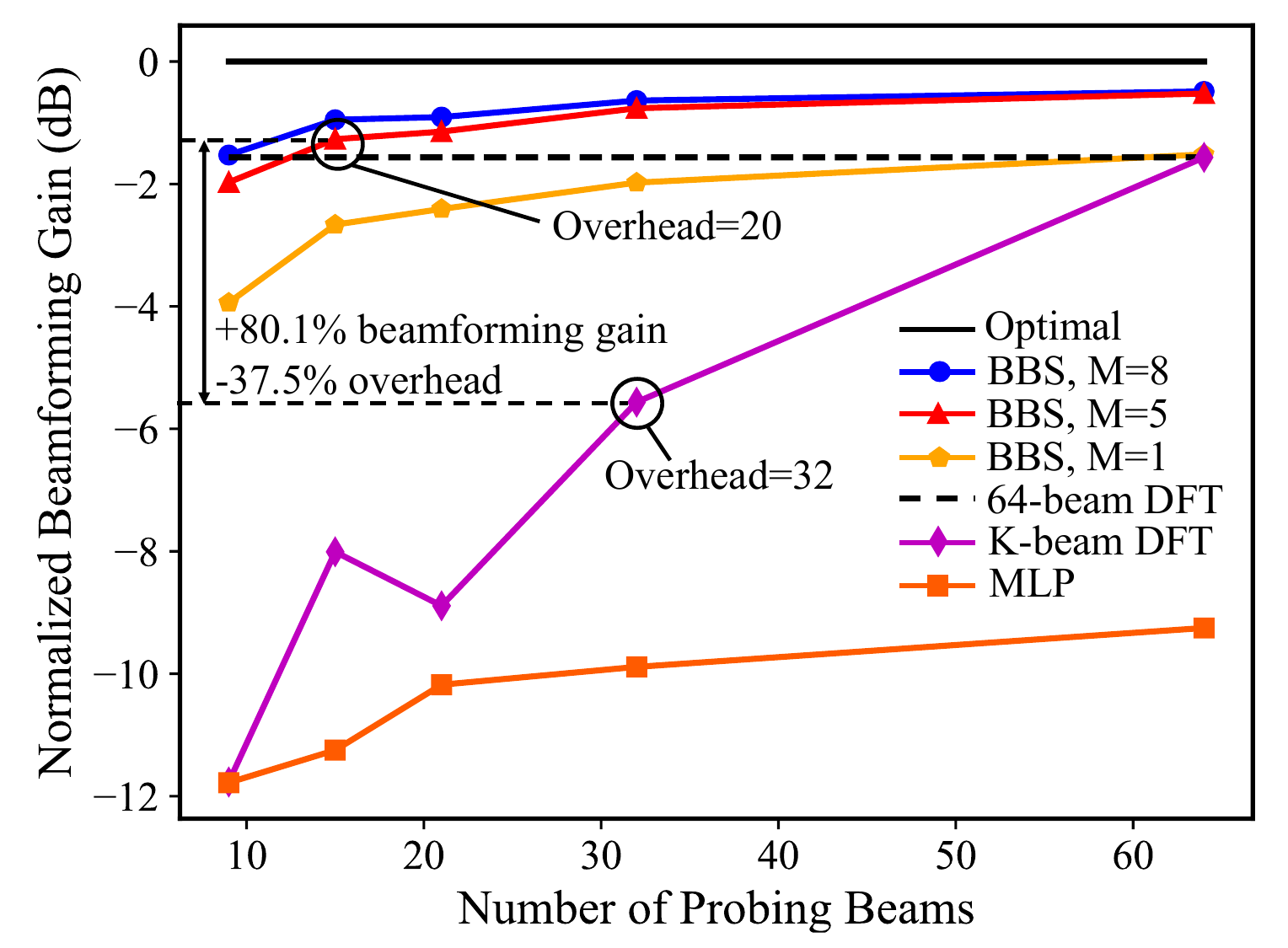}
	\caption{Normalized beamforming gain vs. $Q$ in Boston5G\_28 scenario.}
	\label{figure_11}
\end{figure}

We further compare the generated beam pattern against the optimal one and that obtained from an exhaustive search over a 64-beam DFT codebook. In contrast to conventional DFT beams that typically exhibit a single main lobe, the generated beams often have multiple main lobes. Such beam patterns can likely capture more information about the propagation environment, thereby enhancing the beamforming gain. Taking O1B\_28 scenario as an example as shown in Fig. \ref{figure_7}, due to the blockage in front of the BS and the reflectors placed on its two sides as shown by the geometric in Fig. \ref{figure_5}(b), the generated beam also concentrates most of its energy toward these directions accordingly, which exhibits similar characteristics of the optimal beam.

Moreover, the influence of wireless prompt length on the generation quality has also been investigated. For I2\_28B in Fig. \ref{figure_6}, as the length of RSRP vector increases from 15 to 32, the two main lobes at around $\pm 45\degree$ become sharper, the lobe at $0\degree$ is adjusted to the left by about $10\degree$. Collectively, these changes result in a beam pattern that more closely approximates the optimal one. For O1B\_28 in Fig. \ref{figure_7}, the primary benefit of increasing the RSRP vector length lies in the suppression of side lobes. As shown in Fig. \ref{figure_7}(a) and (c), while their main lobe shapes at $t=1000$ are nearly identical, the case with $Q=32$ achieves a more effective suppression of side lodes near $0\degree$ (In next subsection, we will see that there is a trade-off between prompting length and the achievable beamforming gain). For Boston5G\_28 in fig. \ref{figure_8}, the main difference is the appearance of a main lobe at about $-55\degree$.

\subsection{Performance of Beamforming Gain and Overhead}
While qualitative analysis provides an intuitive assessment of whether the BBS can generate ``meaningful'' beams, quantitative analysis, in turn, offers a clear measurement of the specific gain it delivers. The normalized beamforming gain is considered as the key metric because it is positively correlated with SNR. The normalized beamforming gain can be computed by
\begin{equation}
	\bar{g} = 10{\rm log}_{10}\left(\frac{\vert \bm{{\rm h}}^{\rm H}\bm{{\rm w}} \vert^2}{\vert \bm{{\rm h}}^{\rm H}\bm{{\rm w}}_{{\rm MRT}} \vert^2}\right),
\end{equation}
where $\bm{{\rm w}}_{{\rm MRT}}$ is the beamforming vector using MRT, which is served as the optimal solution. Fig. \ref{figure_9} - \ref{figure_11} demonstrate the normalized beamforming gain versus the length of the wireless prompts under I2\_28B, O1B\_28 and Boston5G\_28 scenarios, respectively. First, it is observed that in all 3 environments, the normalized beamforming gain of the proposed BBS improves with the number of probing beams. This is because more power measurements can provide richer site-specific information for a UE, that is, the wireless prompt is more informative. In addition, compared with the "you-only-think-once (YOTO)" scheme ($M=1$), brainstorming ($M>1$) yield a significant performance improvement. This is because for the trained generative model, its parameters constitute a latent beam space capable of generating (without explicit storage) a virtually unlimited set of high-quality and site-specific beams on-demand for any UE. Consequently, utilizing a set of such customized beams typically leads to a superior performance. For instance, in I2\_28B scenario, compared to the 32-beam DFT scheme, our BBS solution (with $Q=9$ and $M=5$) achieves an 84.85\% improvement in beamforming gain while simultaneously reducing the beam sweeping overhead by 56.25\%. When compared against the exhaustive 64-beam DFT search, the proposed solution offers a 64.74\% gain improvement with a significant overhead reduction of 78.1\%. The complete percentage changes in beam sweeping overhead and the corresponding beamforming gain for the proposed BBS and exhaustive search are summarized in table \ref{table2}, where S1, S2 and S3 represents I2\_28B, O1B\_28 and Boston5G\_28, respectively. Moreover, the beam prediction based approach using discriminative regression is also compared. Nevertheless, it is observed that this scheme yields the lowest average beamforming gain performance across the three scenarios. This underperformance may stem from the model's insufficient understanding of the environments, which could render its adaption counterproductive. Specifically, under the same overhead ($Q=32, M=1$), BBS can achieve an average beamforming gain improvement of 76.3\% in three scenarios.

Furthermore, it is noted that increasing $M$ from 1 to 5 results in a significant beamforming gain improvement, whereas a further increase to 8 offers limited additional gain but incurs higher beam sweeping overhead. It means that there is a trade-off in the intensity of the brainstorming. Similarly, a threshold effect is also observed for the length of wireless prompts $Q$. Exceeding this threshold yields only marginal performance gains while incurring increased overhead. How to optimize these trade-offs is an interesting point, which is left for future work.

\begin{table}
	\centering
	\caption{percentage Change in Beam Sweeping Overhead and Beamforming Gain}
	\label{table2}
	\scalebox{0.9}{
	\begin{tabular}{|c|c|c|c|c|c|c|}
		\hline
		\multirow{6}{*}{S1} & $\Delta O$ (BBS-M1) & -85.9\% & -76.5\% & -67.2\% & -50\% & 0\% \\
		& $\Delta \bar{g}$ (BBS-M1) & -0.3\% & +0.02\% & +0.1\% & +29.3\% & +43.8\% \\
		\cline{2-7}
		& $\Delta O$ (BBS-M5) & -78.1\% & -68.7\% & -59.3\% & -42.1\% & +0.07\% \\
		& $\Delta \bar{g}$ (BBS-M5) & +64.7\% & +69.2\% & +74.2\% & +77.2\% & +85.6\% \\
		\cline{2-7}
		& $\Delta O$ (BBS-M8) & -73.4\% & -64.1\% & -54.7\% & -37.5\% & +12.5\% \\
		& $\Delta \bar{g}$ (BBS-M8) & +72.9\% & +77.5\% & +78.8\% & +84.5\% & +90.3\% \\
		\hline
		\multirow{6}{*}{S2} & $\Delta O$ (BBS-M1) & -85.9\% & -76.5\% & -67.2\% & -50\% & 0\% \\
		& $\Delta \bar{g}$ (BBS-M1) & -19.7\% & -16.5\% & -10.9\% & +0.01\% & +18.2\% \\
		\cline{2-7}
		& $\Delta O$ (BBS-M5) & -78.1\% & -68.7\% & -59.3\% & -42.1\% & +0.07\% \\
		& $\Delta \bar{g}$ (BBS-M5) & +47\% & +53.1\% & +53.1\% & +56.5\% & +70.7\% \\
		\cline{2-7}
		& $\Delta O$ (BBS-M8) & -73.4\% & -64.1\% & -54.7\% & -37.5\% & +12.5\% \\
		& $\Delta \bar{g}$ (BBS-M8) & +55.2\% & +58.3\% & +59.8\% & +62.8\% & +77.8\% \\
		\hline
		\multirow{6}{*}{S3} & $\Delta O$ (BBS-M1) & -85.9\% & -76.5\% & -67.2\% & -50\% & 0\% \\
		& $\Delta \bar{g}$ (BBS-M1) & -151.6\% & -70.1\% & -53.7\% & -26.3\% & +0.03\% \\
		\cline{2-7}
		& $\Delta O$ (BBS-M5) & -78.1\% & -68.7\% & -59.3\% & -42.1\% & +0.07\% \\
		& $\Delta \bar{g}$ (BBS-M5) & -0.25\% & +19\% & +26.8\% & +51.3\% & +66.6\% \\
		\cline{2-7}
		& $\Delta O$ (BBS-M8) & -73.4\% & -64.1\% & -54.7\% & -37.5\% & +12.5\% \\
		& $\Delta \bar{g}$ (BBS-M8) & +0.02\% & +39.3\% & +42\% & +59.2\% & +68.8\% \\
		\hline
	\end{tabular}}
\end{table}

\subsection{The Impact of Noisy Wireless Prompts}
\begin{figure*}[t]
	\centering
	\includegraphics[width=7.0in]{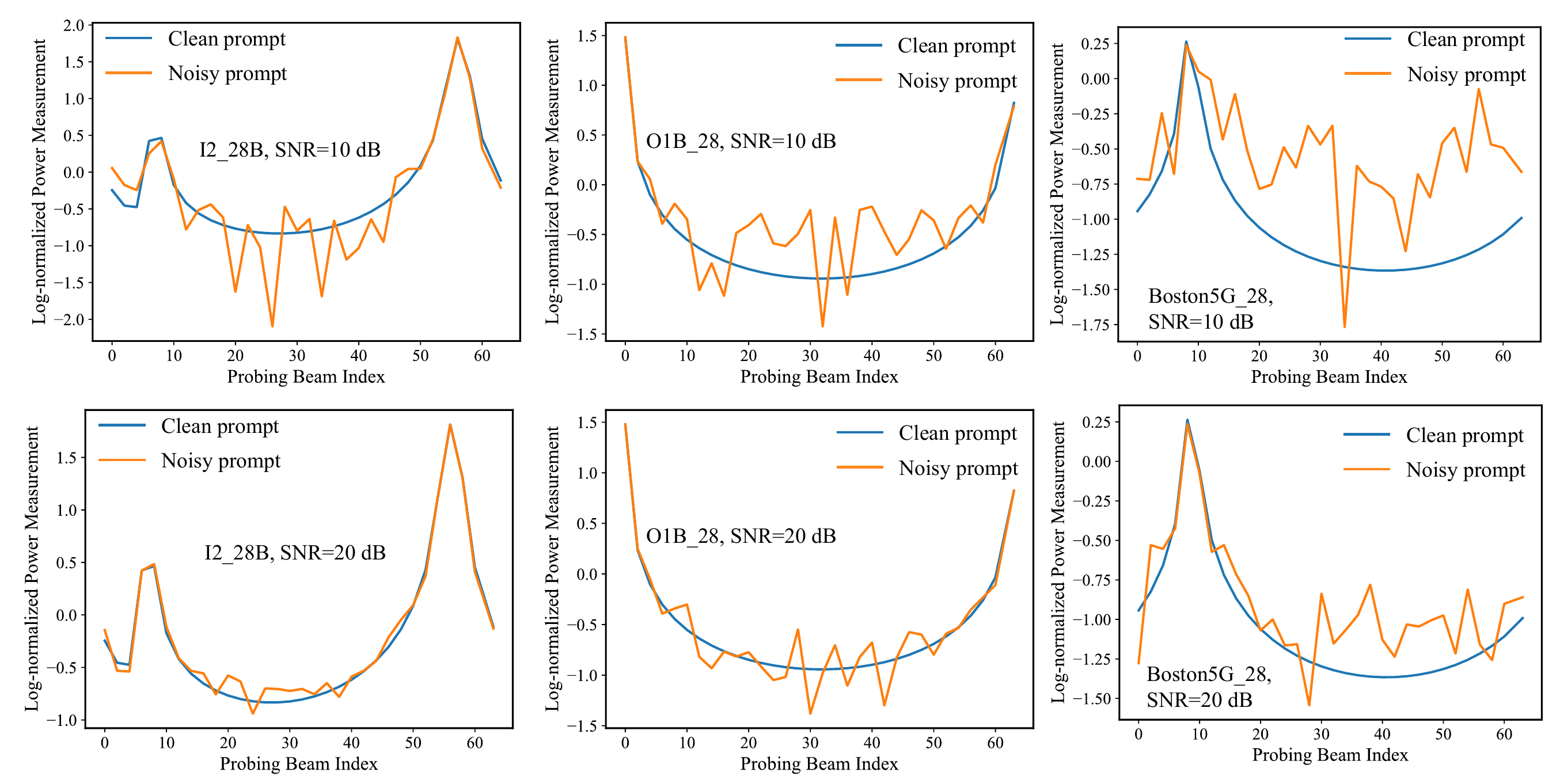}
	\caption{Noisy wireless prompts in different scenarios and SNR.}
	\label{figure_12}
\end{figure*}

\begin{figure}[h]
	\centering
	\includegraphics[width=3.5in]{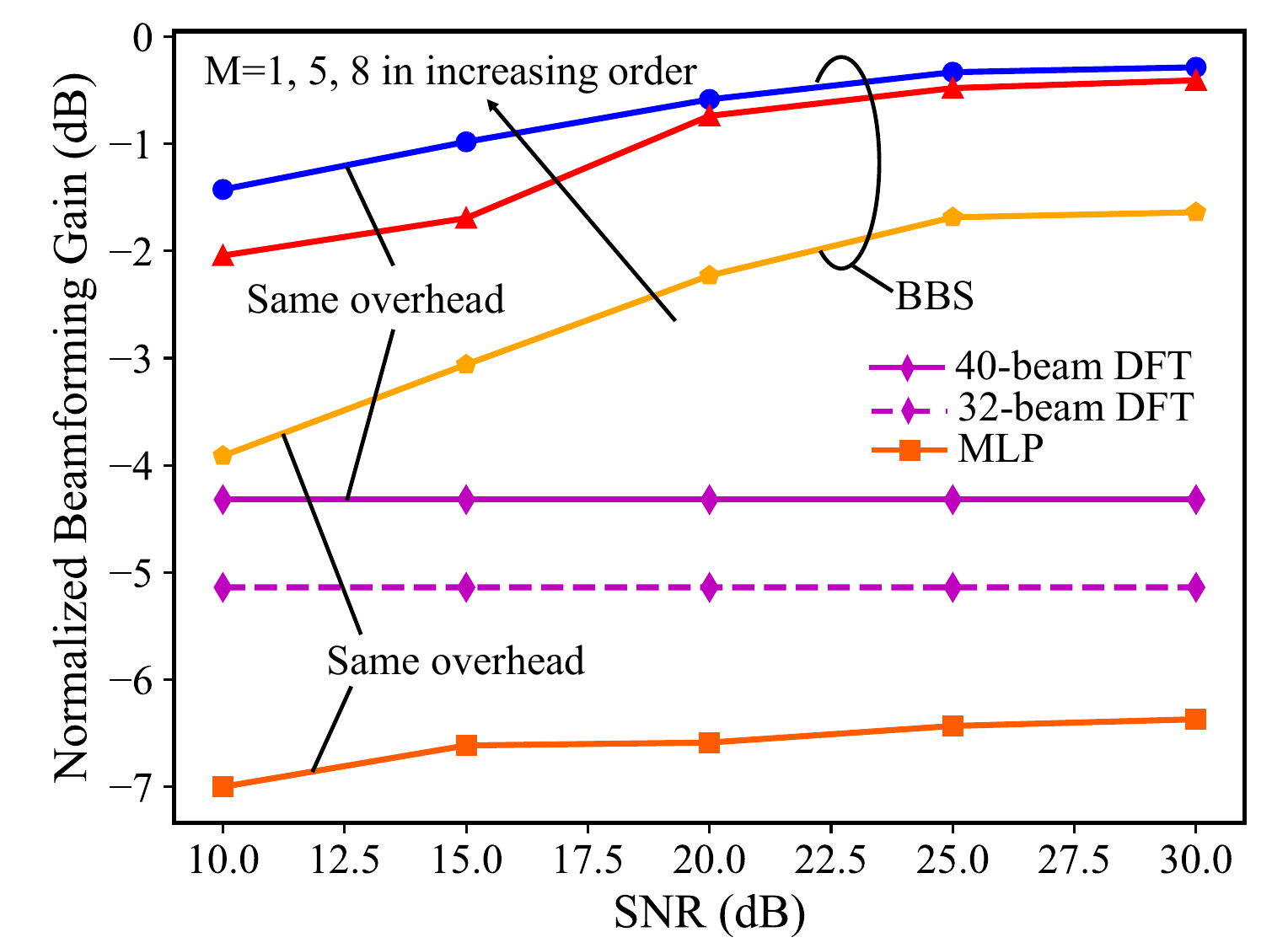}
	\caption{Normalized beamforming gain vs. SNR in I2\_28B scenario.}
	\label{figure_13}
\end{figure}

\begin{figure}[h]
	\centering
	\includegraphics[width=3.5in]{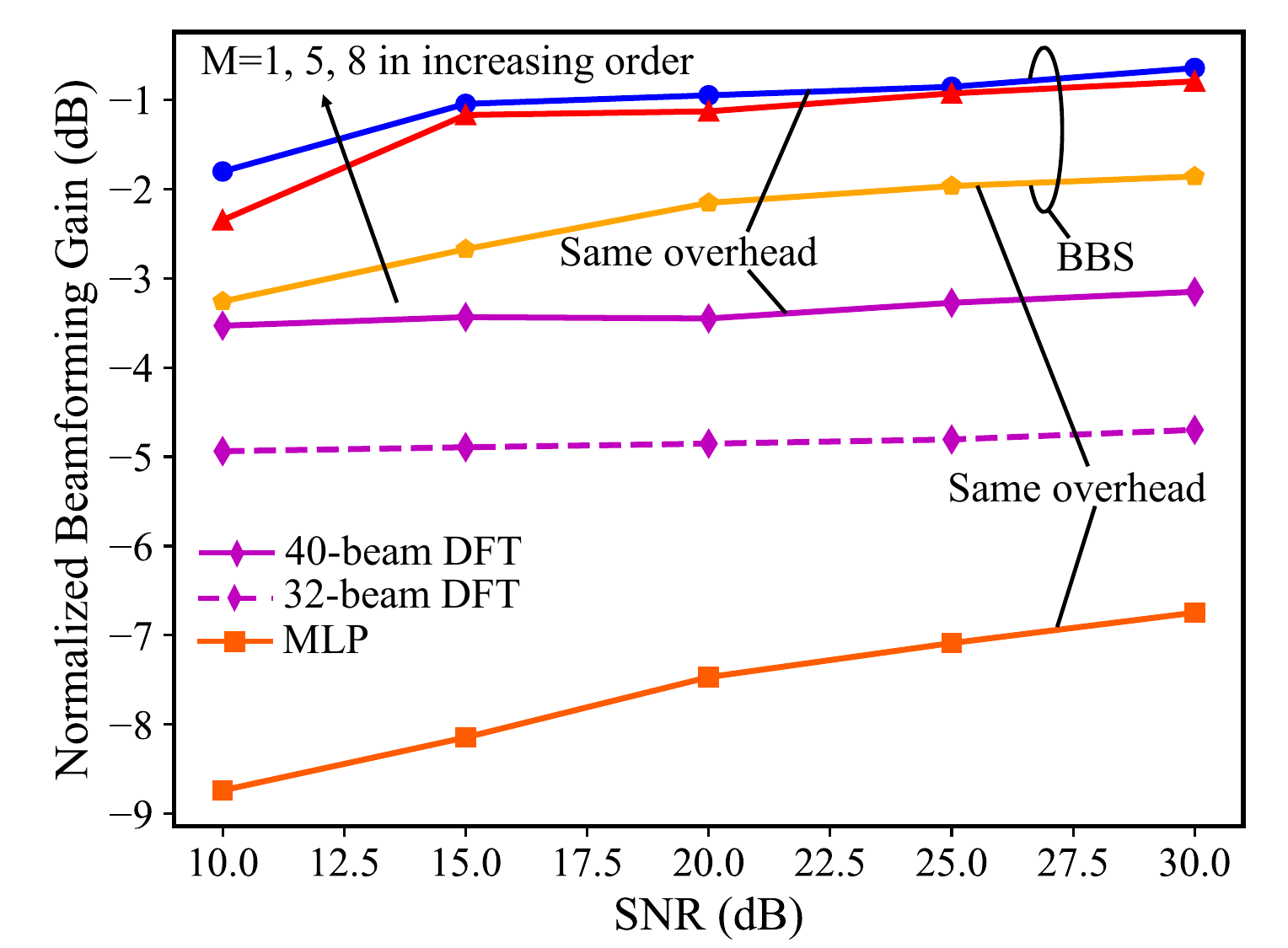}
	\caption{Normalized beamforming gain vs. SNR in O1B\_28 scenario.}
	\label{figure_14}
\end{figure}

\begin{figure}[h]
	\centering
	\includegraphics[width=3.5in]{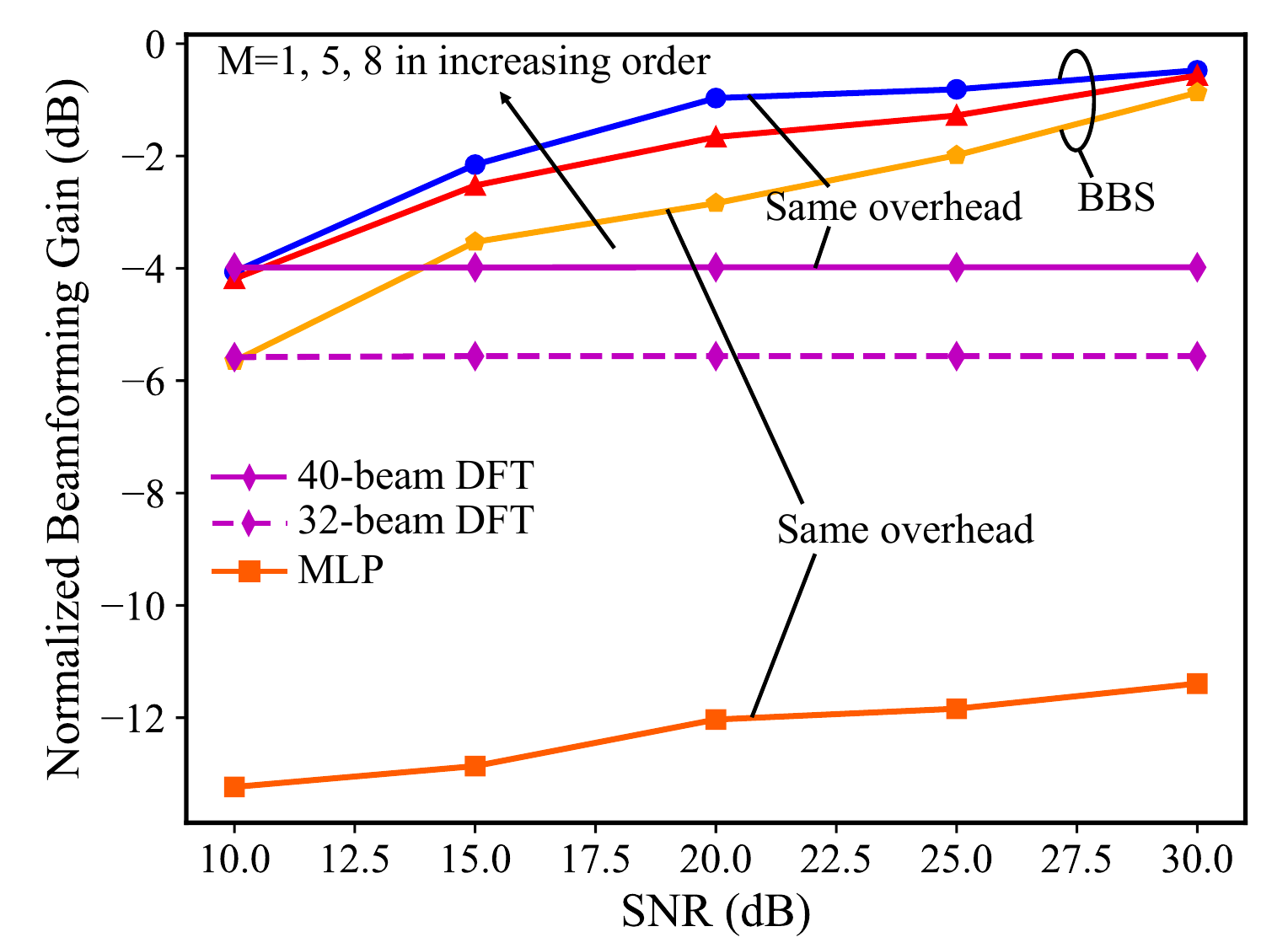}
	\caption{Normalized beamforming gain vs. SNR in Boston5G\_28 scenario.}
	\label{figure_15}
\end{figure}

For evaluating the practicability, it is important to test the performance under noisy environments. In this subsection, the impact of noisy wireless prompts on the GenSSBF performance is evaluated. Specifically, the noise power is computed based on the median received power of the probing DFT-beams and the SNR. The noisy RSRP vectors in three different scenario are illustrated in Fig. \ref{figure_12}, with the SNR changes from 10 dB to 30 dB. It is observed that a higher noise level corresponds to increased fluctuations in RSRP vector, particularly for those with lower power measurements. The normalized beamforming gain at various SNR levels in I2\_28B, O1B\_28 and Boston5G\_28 are shown in Fig. \ref{figure_13}-Fig. \ref{figure_15}, respectively. Obviously, the beamforming gain of BBS drops with decreasing SNR because a more ``dirty'' condition can provide inaccurate guidance. However, thanks to the better understanding of the environments, the proposed BBS can still achieve superior beamforming gain performance in low SNR environment. For instance, in I2\_28B scenario as shown in Fig. \ref{figure_13}, under the same beam sweeping overhead, BBS outperforms the exhaustive searching with beamforming gain improvement of 66.99\% at 10 dB SNR and 93.38\% at 30 dB SNR. However, another important observation is: since the wireless prompts used in BBS is obtained by probing with DFT-beams, thus the robustness of BBS is fundamentally bounded by the DFT codebook. Additionally, the RSRP vector comprises both high- and low power measurements. At low SNR case, especially when the noise power exceeds the received power, the useful information within the vector is dominated by noise, resulting in performance degradation. Therefore, a very important direction of future work is designing more robust wireless prompt structures.

\section{Conclusion}\label{Section5}
This paper proposed a unified design framework for GenSSBF, comprising site profile, wireless prompting module and a generator. As a concrete instantiation of this GenSSBF framework, we further proposed a BBS solution via conditional diffusion model and RSRP prompts, which can be seamless integrated into current 5G systems without changing the standardized protocols. The proposed BBS solution was extensively evaluated using the DeepMIMO datasets, covering both indoor and outdoor scenarios with LoS and NLoS links. Simulation results demonstrated that: 1) mild brainstorming can markedly improve the beamforming performance; 2) compared to the benchmark approaches, BBS significantly reduces the beam sweeping overhead while effectively enhancing the beamforming gain; and 3) even in noisy environments, the proposed BSS still achieves higher beamforming gain, indicating its capability to perform in real-world wireless propagation scenarios. Future work will explore the design of other wireless prompt types with enhanced robustness, and the GenSSBF framework in multi-user and multi-cell settings.

\newpage

\vspace{11pt}

\vfill

\end{document}